\documentclass[journal]{IEEEtran}
\usepackage{graphics} % for pdf, bitmapped graphics files
\usepackage{subfigure}
\usepackage{epsfig} % for postscript graphics files
\usepackage{multicol}
\usepackage{amssymb}  % assumes amsmath package installed
\usepackage{mathtools}
\usepackage{amsmath} % assumes amsmath package installed
\usepackage{mathrsfs}
\usepackage{amsmath,amsfonts,bm}
\usepackage{algorithm}
\usepackage{algorithmicx}
\usepackage{algpseudocode}
\usepackage{epstopdf}
\usepackage{multirow}
\usepackage{comment}
\usepackage{caption}
\usepackage{soul,xcolor}
\usepackage{hyperref}
\usepackage{array}
\usepackage{balance}
\usepackage{threeparttable}
\newtheorem{theorem}{Theorem}
\newtheorem{remark}{Remark}
\newtheorem{lemma}{Lemma}

\newtheorem{definition}{Definition}
\newtheorem{proposition}{Proposition}
\newcommand{\colvec}[2][.8]{%
	\scalebox{#1}{%
		\renewcommand{\arraystretch}{.8}%
		$\begin{bmatrix}#2\end{bmatrix}$%
	}
}

\newcommand{\method}[1]{\texttt{#1}}
\usepackage{cite}

\ifCLASSINFOpdf
\else
\fi
\usepackage{amsmath}
\usepackage{array}
\begin{document}
	\title{Privacy-Preserving Data-Enabled Predictive Leading Cruise Control in Mixed Traffic}
	
	\author{
		Kaixiang~Zhang, Kaian~Chen,
		Zhaojian~Li, Jun~Chen, and Yang~Zheng 
		\thanks{This work was supported in part by the National Science Foundation (NSF) under Grant 2045436, Grant CMMI-2320698, and Grant CMMI-2320697. The work of Yang Zheng was also in part supported by NSF under Grant ECCS-2154650. \textit{(Kaixiang Zhang and Kaian Chen contributed to this work equally.)} \textit{(Corresponding author: Zhaojian Li.)}}
		\thanks{Kaixiang Zhang, Kaian Chen, and Zhaojian Li are with the Department of Mechanical Engineering, Michigan State University, East Lansing, MI 48824, USA (e-mail: \{zhangk64,chenkaia,lizhaoj1\}@msu.edu).}
		\thanks{Jun Chen is with the Department of Electrical and Computer Engineering, Oakland University, Rochester, MI 48309, USA (email: junchen@oakland.edu).}
		\thanks{Yang Zheng is with the Department of Electrical and Computer Engineering, University of California San Diego, CA 92093, USA, (e-mail: zhengy@ucsd.edu).}
	}
	% The paper headers
	\markboth{}%
	{Shell \MakeLowercase{\textit{et al.}}: Bare Demo of IEEEtran.cls for IEEE Journals}
	
	% make the title area
	\maketitle

\begin{abstract}
	Data-driven predictive control of connected and automated vehicles (CAVs) has received increasing attention as it can achieve safe and optimal control without relying on explicit dynamical models. However, employing the data-driven strategy involves the collection and sharing of privacy-sensitive vehicle information, which is vulnerable to privacy leakage and might further lead to malicious activities. In this paper, we develop a privacy-preserving data-enabled predictive control scheme for CAVs in a mixed traffic environment, where human-driven vehicles (HDVs) and CAVs coexist. We tackle external eavesdroppers and honest-but-curious central unit eavesdroppers who wiretap the communication channel of the mixed traffic system and intend to infer the CAVs' state and input information. An affine masking-based privacy protection method is designed to conceal the true state and input signals, and an extended form of the data-enabled predictive leading cruise control under different data matrix structures is derived to achieve privacy-preserving optimal control for CAVs. Numerical simulations demonstrate that the proposed scheme can protect the privacy of CAVs against~attackers without affecting control performance or incurring heavy computations. %Numerical simulations demonstrate the efficacy of the developed approach.
\end{abstract}

	
\IEEEpeerreviewmaketitle
	
\section{Introduction}
	
\IEEEPARstart{A}{dvances} in wireless communication technologies such as vehicle-to-infrastructure (V2I) or vehicle-to-vehicle (V2V) offer modern vehicles with enhanced connectivity and new opportunities for intelligent and integrated vehicle control~\cite{Vahidi2018ET,Li2017ITSM}. One typical example is the cooperative adaptive cruise control (CACC), which regulates a convoy of vehicles to improve traffic flow stability, safety, and energy efficiency~\cite{Milanes2013TITS,Zheng2017TCST,Wang2020TR,Guo2020ITJ}. While attracting much research interest, existing studies on CACC predominately focus on effective platooning of full connected and automated vehicles (CAVs). However, due to the gradual deployment of CAVs, it is expected that human-driven vehicles (HDVs) and CAVs will coexist for a long period of time, which raises the demand for developing cooperative control of CAVs in a mixed traffic flow.
	
HDVs often exhibit stochastic and uncertain behaviors that are difficult to characterize, making it challenging for the control design of CAVs in a mixed traffic environment. To address this issue, several approaches have been proposed. For instance,  feedback controllers of CAVs are designed under the formation of connected cruise control in~\cite{Orosz2016VSD,Jin2016TITS,Jin2018TR}, where CAVs can receive information from HDVs ahead. Leading cruise control (LCC) is proposed in~\cite{Wang2021TITS}, where CAVs make decisions by utilizing both the preceding and following HDVs' information. To handle the prediction uncertainty of HDVs, a robust platoon control framework is designed in~\cite{Feng2021TIV} by leveraging tube model predictive control (MPC). We note that existing studies mainly focus on designing model-based approaches for CAVs~\cite{Wang2021TITS,Orosz2016VSD,Jin2016TITS,Jin2018TR,Feng2021TIV,Zheng2020ITJ,Wang2020TITS}. One common strategy is to utilize car-following models to characterize the behavior of human drivers, e.g., optimal velocity model (OVM)~\cite{Bando1995PRE} or intelligent driver model~\cite{Treiber2000PRE}, enabling a state-space model of the mixed traffic system for system analysis and control design. Although model-based approaches can provide rigorous theoretical analysis and control synthesis when an accurate traffic model is available, it might not be applicable to real-world deployment since the parameters in human car-following models are non-trivial to identify accurately.
	
Instead of relying on explicit system models, data-driven approaches have recently emerged as an alternative to avoid model/parameter identification and directly incorporate collected data for control designs. For example, reinforcement learning (RL)~\cite{Chu2019CDC,Vinitsky2018ITSC,Chen2021arXiv} and adaptive dynamic programming (ADP)~\cite{Gao2016TITS,Huang2020TCYB,Liu2021CDC} have been developed to design optimal control schemes for CAVs. Although RL-based techniques can use experimental data to learn a model to simulate uncertain HDV behaviors, they are typically data intensive and have limited interpretability. On the other hand, data-driven ADP algorithms can provide optimal control actions, %with rigorous stability guarantees, 
but they struggle to handle constraints that are critical to vehicle safety. Recent advances in data-driven MPC have shown promise to achieve optimal control with constraint and stability guarantees~\cite{Berberich2020TAC,Coulson2021TAC}. In~\cite{Wang2022arXiv,wang2022experimental}, a Data-EnablEd Predictive Leading Cruise Control (\method{DeeP-LCC}) strategy is developed for a mixed traffic system, which can efficiently handle safety constraints among multiple CAVs and HDVs. Specifically, by leveraging the Data-EnablEd Predictive Control (\method{DeePC}) techniques~\cite{Coulson2019ECC}, input/output measurements are collected to first represent the mixed traffic system in a non-parametric manner, and to subsequently formulate a constrained optimization problem~\cite{Wang2022arXiv}. \method{DeePC} exploits behavioral system theory~\cite{Markovsky2021ARC} and Willem's fundamental lemma~\cite{Willems2005SCL} to implicitly describe the system trajectories without explicitly carrying out model identification. Its receding horizon implementation is shown to be equivalent to the MPC formulation for  linear time-invariant (LTI) systems, and it has found various successes in several practical applications~\cite{Huang2021TCST,Elokda2021IJRNC}.
	
Employing the \method{DeeP-LCC} approach for the control of CAVs in the mixed traffic, however, poses great concerns in privacy. To enable coordinated control, the onboard data of vehicles, which may contain private information, needs to be extensively collected and shared via wireless V2I or V2V communications, % (e.g., V2I or V2V), 
causing potential privacy leakage. Specifically, in a typical control architecture, % for a mixed-traffic fleet, 
each vehicle (both HDVs and CAVs) first sends its measured/estimated states to a central unit (e.g., a road-side edge or a remote cloud). The central unit then solves a pre-specified mixed-traffic optimal control problem and sends optimal control actions back to CAVs. In this setup, system measurements and control actions of CAVs need to be transmitted between the central unit and the local vehicles, raising concerns that an external eavesdropper or an honest-but-curious central unit can wiretap the communication channels to get sensitive information. In fact, several studies have shown that exposing local vehicle's information to connectivity can lead to security vulnerabilities and various malicious activities~\cite{PetitTITS2015,Amoozadeh2015CM,Sun2022TITS}. Failing to protect privacy can potentially lead to devastating effects for CAVs and other vehicles sharing the roadway.

The growing awareness of security in cyber-physical systems makes it imperative to protect privacy for CAV control in mixed traffic. So far, privacy and security problems have been studied under various intelligent vehicle scenarios~\cite{Farivar2021TITS,Biroon2020ACC,Gao2022TITS}. In particular, numerous privacy-preserving approaches have been developed to enhance communication security of intelligent vehicles by leveraging conventional information technology privacy mechanisms, e.g., cryption~\cite{Jarouf2022IVS,Kamal2023TITS}, secret sharing~\cite{Li2019}, and differential privacy~\cite{Ghane2019,Ghane2021TITS}. However, these approaches are not appropriate for the mixed traffic control considered in this work for two reasons. First, most of them are designed to protect the privacy of vehicles' information against the external eavesdropper, but cannot tackle honest-but-curious adversary which follows all communication/computation protocols correctly but is curious and uses received messages to infer vehicles' private information. Second, conventional privacy mechanisms either trade accuracy for privacy (e.g., differential privacy) or incur heavy computation/communication overhead (e.g., cryption), and hence are inappropriate for mixed traffic system subject to stringent accuracy and real-time constraints.

In this paper, we develop the first privacy-preserving data-enabled predictive control scheme for controlling CAVs in mixed traffic. Specifically, we consider the same mixed traffic system in~\cite{Wang2022arXiv}, where multiple CAVs and HDVs cooperate with each other under the LCC framework~\cite{Wang2021TITS}. We show that if the central unit is an honest-but-curious adversary or there exists an external eavesdropper, the \method{DeeP-LCC} under both Hankel and Page matrix structures cannot protect the private information of the vehicles. To avoid leaking the state and input information of the CAVs, a simple yet effective affine masking-based privacy protection method is designed, which can mask the true state and input signals. After the affine masking, an extended \method{DeeP-LCC} is derived to generate safe and optimal control actions for the CAVs. We further introduce a privacy notion and show that the proposed affine masking-based method can protect the private system state and input signals from being inferred by the attackers. Some preliminary results are summarized in a conference version~\cite{Zhang2022ITSC}. The new contributions of this work are as follows.
	
First, we study non-parametric representations of the mixed traffic system under both Hankel and Page matrix structures. The previous \method{DeeP-LCC} scheme \cite{Wang2022arXiv} only leverages the Hankel matrix to store the collected data for non-parametric system representation. We extend \method{DeeP-LCC} with the Page matrix which is another effective structure for time-series data \cite{Damen1982SCL}. Second, we propose a privacy-preserving \method{DeeP-LCC} method for mixed traffic control, which can retain the advantages of the original \method{DeeP-LCC} while avoiding privacy leakage. Specifically, we conceal the privacy-sensitive state and input signals of CAVs via affine masking and reformulate a compatible \method{DeeP-LCC} that is equivalent to the original one. Although affine masking has been utilized for privacy protection in cloud-based MPC~\cite{Sultangazin2020TAC,Zhang2021privacy}, its application in data-driven approaches has not been thoroughly explored. The challenge arises from the fundamental differences between data-driven control and MPC in problem formulation and theoretical analysis. In this paper, we successfully incorporate affine masking with data-driven predictive control to enhance the privacy protection of the mixed traffic system. This newly developed \method{DeeP-LCC} is a non-trivial extension that ends with a new \method{DeePC} structure and provable privacy guarantees. The affine masking-based method is light-weight in computation, making it suitable for CAV controls that have fast dynamics. Finally, we conduct extensive traffic simulations to validate the performance of the privacy-preserving \method{DeeP-LCC}. The results demonstrate the benefits of our proposed method in terms of not only improving traffic smoothness and fuel economy, but also protecting privacy against potential attackers.
	
The remainder of this paper is organized as follows.~Section~\ref{sec_formulation} introduces background notations and formulates the mixed traffic control problem.  Section~\ref{sec_DeeP} provides an overview of \method{DeeP-LCC} method from~\cite{Wang2022arXiv}. Section~\ref{sec_privacyPreserved} presents the affine masking-based privacy protection method and the extended \method{DeeP-LCC}. Section~\ref{sec_privacyAnalysis} analyses the equivalence and privacy-preserving properties of the developed method. Simulations are shown in Section~\ref{sec_perfEva}. Finally, we conclude the paper in Section~\ref{sec_conclusion}. 

\begin{figure*}[t]
	\centering
	\includegraphics[width=0.8\textwidth]{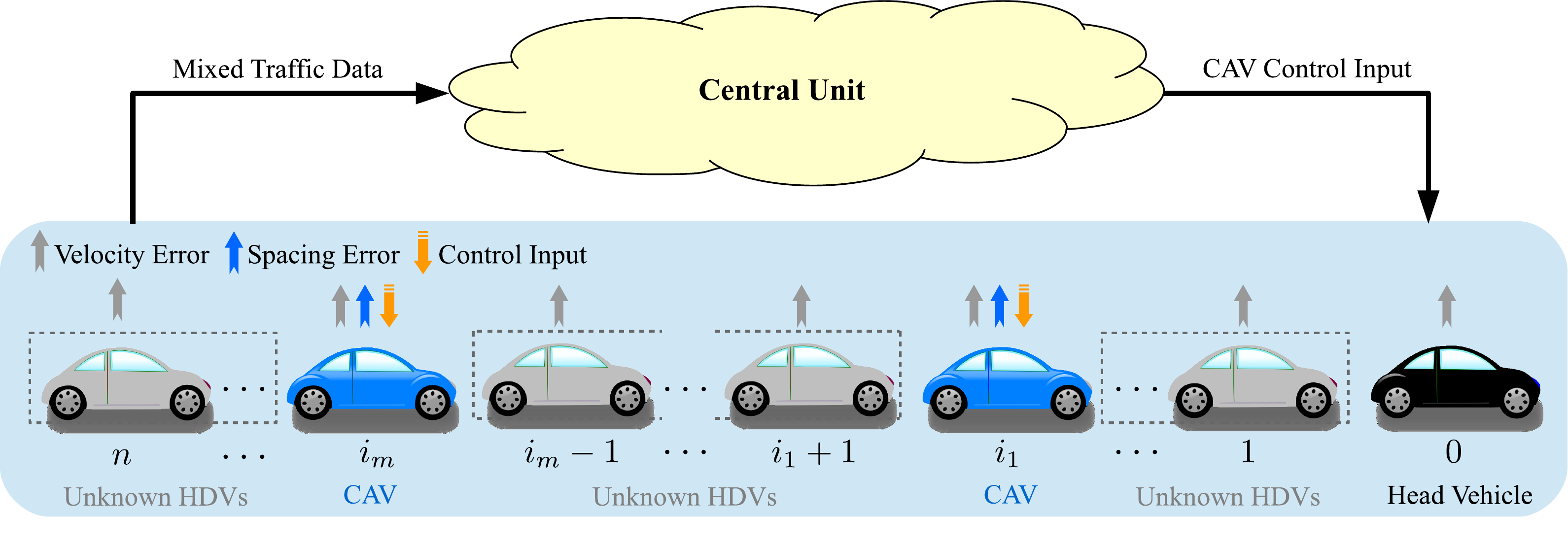}
	\caption{Schematic of mixed traffic system with $n+1$ vehicles.}
	\label{Fig:SystemSchematic}
\end{figure*}
	
\textit{Notations:} We denote $\mathbb{R}$ and $\mathbb{Z}_{+}$ as the set of real numbers and positive integers, respectively. $0_{n}$ and $0_{n\times m}$ are denoted as a zero vector of size $n$ and a zero matrix of size $n\times m$, respectively. $1_n$ and $I_{n}$ are denoted as an $n\times 1$ vector with all entries being ones and an $n\times n$ identity matrix, respectively. 
Denote $\mathrm{diag}(a_{1},\ldots,a_{n})$ as a diagonal matrix with $a_{1}, \ldots, a_{n}$ on its diagonal entries, and $\mathrm{diag}(A_{1},\ldots,A_{n})$ as a block-diagonal matrix with matrices $A_{1},\ldots,A_{n}$ on its diagonal blocks. $e_{n}^m$ is used to denote an $n \times 1$ vector, with the $m$-th entry being one and the others being zero.
	
\section{Modeling and Problem Statement} \label{sec_formulation}
	
\subsection{System Model of Mixed Traffic} \label{subsec_systemModel}
As shown in Fig.~\ref{Fig:SystemSchematic}, we consider a mixed traffic system consisting of $n+1$ vehicles: one head vehicle (indexed as $0$), $m$ CAVs, and $n-m$ HDVs. The head vehicle is controlled by a human driver. % and is essentially the HDV ahead of the first CAV. 
Let $\Omega=\{1,2,\ldots,n\}$ be the set of vehicle indices ordered from front to end. The sets of CAV indices and HDV indices are denoted by $\Omega_{C}=\{i_1,i_2,\ldots,i_m\}\subseteq \Omega$ and $\Omega_{H}=\{j_1,j_2,\ldots,j_{n-m}\}= \Omega \backslash \Omega_{C}$, respectively, where $i_1 < i_2 < \ldots < i_m$ and $j_1 < j_2 < \ldots < j_{n-m}$. We use $p_i(t)$, $v_i(t)$ and $a_i(t)$ to denote the position, velocity and acceleration of the $i$-th vehicle at time $t$, respectively. 
	
The car-following dynamics for HDVs are modeled by the following nonlinear processes:
\begin{equation}\label{Eq:HDVModel}
	\dot{v}_i(t)=F\left(s_i(t),\dot{s}_i(t),v_i(t)\right), \quad i \in \Omega_{H},
\end{equation}
where $s_i(t)=p_{i-1}(t)-p_i(t)$ and $\dot{s}_i(t)=v_{i-1}(t)-v_i(t)$ are the relative spacing and velocity between vehicle $i$ and its preceding vehicle $i-1$, respectively. Denote $s^{*}$ and $v^{*}$ as the equilibrium spacing and equilibrium velocity of the mixed traffic system. Then the spacing error and velocity error of vehicle $i$ are defined as $\tilde{s}_{i}(t) = s_i(t)-s^*$ and $\tilde{v}_i(t)=v_i(t)-v^*$, respectively. By applying the first-order Taylor expansion on \eqref{Eq:HDVModel} at the equilibrium $\left(s^{*}, v^{*}\right)$, the linearized model for each HDV can be obtained
\begin{equation}\label{Eq:LinearHDVModel}
\begin{cases}
	\dot{\tilde{s}}_i(t)=\tilde{v}_{i-1}(t)-\tilde{v}_i(t),\\
	\dot{\tilde{v}}_i(t)=\alpha_{1}\tilde{s}_i(t)-\alpha_{2}\tilde{v}_i(t)+\alpha_{3}\tilde{v}_{i-1}(t),\\
\end{cases} \, i \in \Omega_{H},
\end{equation}
with $\alpha_{1} = \frac{\partial F}{\partial s}, \alpha_{2} = \frac{\partial F}{\partial \dot{s}} - \frac{\partial F}{\partial v}, \alpha_{3} = \frac{\partial F}{\partial \dot{s}}$ evaluated at the equilibrium state $\left(s^{*}, v^{*}\right)$. In this paper, we use the OVM~\cite{Bando1995PRE} model to describe the HDVs' car-following dynamics in \eqref{Eq:HDVModel}.
More details about the linearized model of OVM model can be found in \cite{Zheng2020ITJ,Wang2020TITS}. For the CAV, the acceleration is used as the control input, i.e., $\dot{v}_i(t)=u_i(t), \, i \in \Omega_{C}$, and its system model is described by
\begin{equation} \label{Eq:LinearCAVModel}
	\begin{cases}
		\dot{\tilde{s}}_i(t)=\tilde{v}_{i-1}(t)-\tilde{v}_i(t),\\
		\dot{\tilde{v}}_i(t)=u_i(t),\\
	\end{cases} \quad i\in \Omega_{C}.
\end{equation}
	
The state of each vehicle is denoted by $x_{i}(t)=\begin{bmatrix}
	\tilde{s}_{i}, \tilde{v}_{i}
\end{bmatrix}^{\top}$, and the overall state of the mixed traffic system is defined as 
\[
x(t) = \begin{bmatrix}
	x_{1}^{\top}(t), x_{2}^{\top}(t), \cdots, x_{n}^{\top}(t)
\end{bmatrix}^{\top} \in \mathbb{R}^{2n}.
\] 
Based on \eqref{Eq:LinearHDVModel} and \eqref{Eq:LinearCAVModel}, the linearized state-space model for the mixed traffic system can be derived, as follows:
\begin{equation} \label{Eq:LinearSystemModel}
	\dot{x}(t)=Ax(t)+Bu(t)+H\epsilon(t),
\end{equation}
where $u(t)=\begin{bmatrix}
	u_{i_1}(t), u_{i_2}(t), \cdots, u_{i_m}(t)
\end{bmatrix}^{\top}$ is the collective control input and $\epsilon(t)=\tilde{v}_{0}(t)=v_0(t)-v^*$ is the velocity error of the head vehicle. The system matrices in~\eqref{Eq:LinearSystemModel} are 
\begin{align*}
A&=\begin{bmatrix} A_{1,1} & & & &   \\
	A_{2,2} & A_{2,1} & &  &   \\
	& \ddots& \ddots&  &  \\
	& & A_{n-1,2}& A_{n-1,1} &  \\
	& & &  A_{n,2}&A_{n,1}\\
\end{bmatrix} \in \mathbb{R}^{2n\times 2n} ,\\
B &= \begin{bmatrix}
	e_{2n}^{2i_{1}}, e_{2n}^{2i_{2}}, \ldots, e_{2n}^{2i_{m}}
\end{bmatrix}\in \mathbb{R}^{2n \times m},\\
H &= \begin{bmatrix}
	1, \alpha_{3}, 0_{2n-2}^{\top}
\end{bmatrix}^{\top}\in \mathbb{R}^{2n},
\end{align*}
where
\begin{align*}
\begin{cases}
	A_{i,1} = 	\begin{bmatrix} 0 & -1 \\ \alpha_{1} & -\alpha_{2} \end{bmatrix}, \; A_{i,2} = \begin{bmatrix} 0 & 1 \\ 0 & \alpha_{3} \end{bmatrix}, \; i\in \Omega_{H};\\
	A_{i,1} = \begin{bmatrix} 0 & -1 \\ 0 & 0 \end{bmatrix}, \; A_{i,2} = \begin{bmatrix} 0 & 1 \\ 0 & 0 \end{bmatrix}, \; i\in \Omega_{C}.
\end{cases}
\end{align*}
	
Recent works on mixed traffic control~\cite{Jin2016TITS,Wang2021TITS,Zheng2020ITJ,Wang2020TITS,Gao2016TITS,Huang2020TCYB,Liu2021CDC} require that both the spacing error and velocity error of HDVs are available. However, as discussed in~\cite{Wang2022arXiv}, the spacing error of HDVs (i.e., $\tilde{s}_i$ $(i \in \Omega_{H})$) is impractical to be observed due to the unknown human driving behavior. Therefore, the output signal of the mixed traffic system is constructed as 
\begin{equation*} %\label{Eq:SystemOutput}
y(t)=\begin{bmatrix}
	x_{i_{1}}^{\top}(t),\ldots, x_{i_{m}}^{\top}(t),
	\tilde{v}_{j_{1}}(t),\ldots,\tilde{v}_{j_{n-m}}(t)
\end{bmatrix}^{\top},
\end{equation*}
where $y(t) \in \mathbb{R}^{n+m}$ consists of all state measurements of the CAVs, i.e., $\tilde{s}_{i}, \tilde{v}_i$ $(i \in \Omega_{C})$, and the velocity error signal of the HDVs, i.e., $\tilde{v}_i$ $(i \in \Omega_{H})$. The output $y(t)$ can be related to the overall state $x(t)$ via
\begin{equation} \label{Eq:SystemOutput}
	y(t)=Cx(t),
\end{equation}
with
\[
C\!=\!\begin{bmatrix}
	e_{2n}^{2i_{1}-1}, e_{2n}^{2i_{1}}, \ldots, e_{2n}^{2i_{m}-1}, e_{2n}^{2i_{m}}, e_{2n}^{2j_{1}},e_{2n}^{2j_{2}},\ldots,e_{2n}^{2j_{n-m}}
\end{bmatrix}^\top
\] being the output matrix. Given the sampling interval $\Delta t > 0$, the continuous-time model in \eqref{Eq:LinearSystemModel} and \eqref{Eq:SystemOutput} can be transformed into the discrete-time
\begin{equation} \label{Eq:DT_TrafficModel}
	\begin{cases}
		x(k+1) = A_\mathrm{d}x(k) + B_\mathrm{d}u(k) + H_\mathrm{d} \epsilon(k),\\
		y(k) = C_\mathrm{d}x(k),
	\end{cases}
\end{equation}
where $
A_\mathrm{d} = e^{A\Delta t}\in \mathbb{R}^{2n\times 2n},B_\mathrm{d}=\int_{0}^{\Delta t} e^{A t} Bd t\in \mathbb{R}^{2n\times m},H_\mathrm{d}=\int_{0}^{\Delta t} e^{A t} Hd t \in \mathbb{R}^{2n},C_\mathrm{d}=C \in \mathbb{R}^{(n+m)\times 2n} 
$. Let $\hat{u}(k)=\begin{bmatrix}
	\epsilon(k), u^{\top}(k)
\end{bmatrix}^{\top}$ be the combined inputs signal and $\hat{B}_{d} = \begin{bmatrix}
	H_{d}, B_{d}
\end{bmatrix}$ be the combined input matrix. Then \eqref{Eq:DT_TrafficModel} can be rewritten into a compact form as follows:
\begin{equation} \label{Eq:DT_TrafficModel2}
	\begin{cases}
		x(k+1) = A_\mathrm{d}x(k) + \hat{B}_\mathrm{d}\hat{u}(k),\\
		y(k) = C_\mathrm{d}x(k).
	\end{cases}
\end{equation}
	
In this paper, we assume that $\alpha_{1}-\alpha_{2}\alpha_{3}+\alpha_{3}^{2} \neq 0$ in \eqref{Eq:LinearHDVModel}, which ensures that the mixed traffic system is stabilizable and observable \cite{Wang2021TITS}. Note that the above model is only used for analysis, and the controller design in Section \ref{sec_DeeP} does not assume the exact model \eqref{Eq:DT_TrafficModel}.

\subsection{Problem Statement}
As shown in Fig.~\ref{Fig:SystemSchematic}, the mixed traffic system utilizes a central unit (e.g., a road-side edge or a remote cloud) to receive all vehicle data and generate the control signals for CAVs. Specifically, each CAV needs to send its state $x_{i}(k)=\begin{bmatrix}
	\tilde{s}_{i}(k), \tilde{v}_{i}(k)
\end{bmatrix}^{\top}$ ($i\in \Omega_{C}$) to the central unit, while the head vehicle and HDVs are requested to provide the velocity error $\epsilon(k)$ and $\tilde{v}_{i}(k)$ ($i\in \Omega_{H}$), respectively. Then, the central unit solves a pre-specified control problem and sends control inputs $u_{i}(k)$ ($i\in \Omega_{C}$) to the CAVs. 

For mixed traffic control, one challenge lies in the unknown HDVs' car-following dynamics, which are difficult to identify due to stochastic and uncertain human behaviors. In addition, the mixed traffic system requires extensive data transmission between vehicles and central unit, which raises concerns of privacy leakage.
Therefore, the main objective of this paper is to design a privacy-preserving data-driven control scheme, which can improve the traffic efficiency of the vehicle fleet without relying on an explicit system model and protect private vehicles' information against attackers. The design details are presented in Sections \ref{sec_DeeP} and \ref{sec_privacyPreserved}.

\section{Data-Enabled Predictive Leading Cruise Control} \label{sec_DeeP}
In this section, we first discuss the non-parametric representations of the mixed traffic system based on Hankel and Page matrix structures and then briefly review the \method{DeeP-LCC} scheme from~\cite{Wang2022arXiv}. 
	
\subsection{Non-Parametric Representation of Mixed Traffic}
Conventional methods rely on explicit system model \eqref{Eq:DT_TrafficModel} to facilitate controller design. One typical example is MPC whose performance  depends closely on the model accuracy. Even though there exist many system identification techniques, it is still difficult to obtain accurate models for complex systems such as  the mixed traffic system with stochastic and uncertain human driving behavior. The \method{DeePC} \cite{Coulson2019ECC} is a promising model-free paradigm and recently has been leveraged by \cite{Wang2022arXiv} to design the \method{DeeP-LCC} scheme for mixed traffic. In particular, based on Willems’ fundamental lemma \cite{Willems2005SCL}, this class of data-driven approaches can describe the dynamical system in a non-parametric manner.

\begin{figure*}[!t]
	\setlength{\abovecaptionskip}{0pt}
	\centering
	\subfigure[] {\label{fig_proposition_p1}
		\includegraphics[width=0.3414\textwidth]{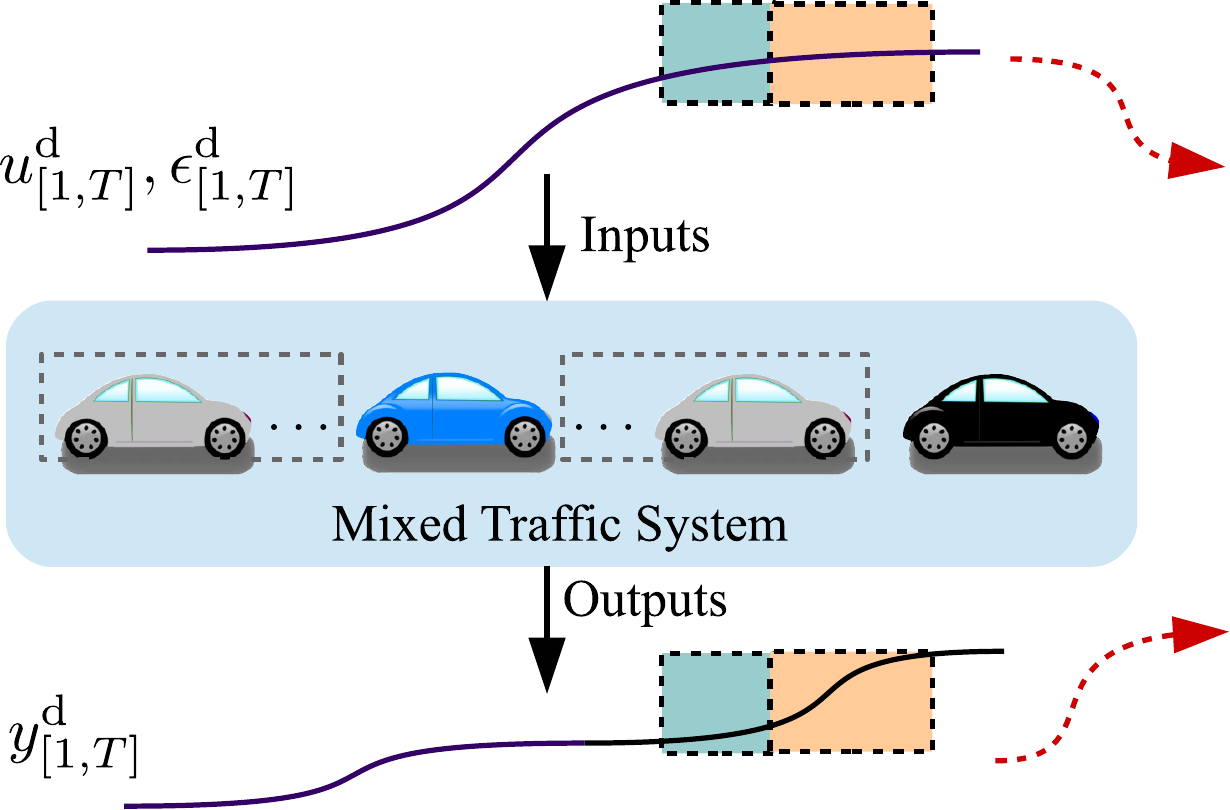}
	}
	\hspace{-3mm}
	\subfigure[] {\label{fig_proposition_p2}
		\includegraphics[width=0.2387\textwidth]{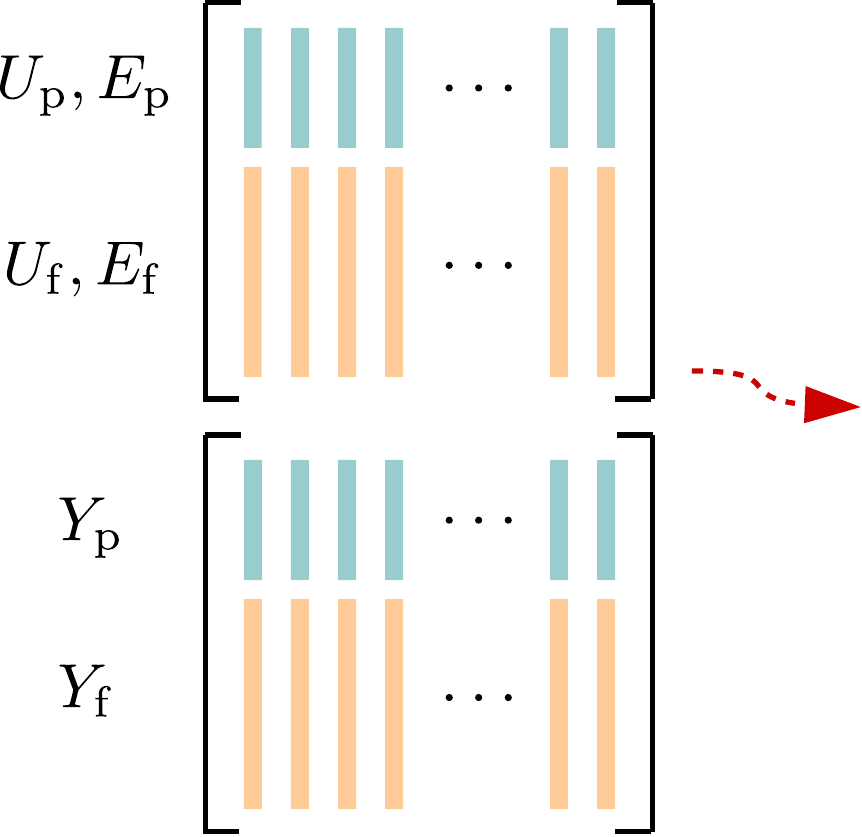}
	}
	\hspace{-3mm}
	\subfigure[]{\label{fig_proposition_p3}
		\includegraphics[width=0.3699\textwidth]{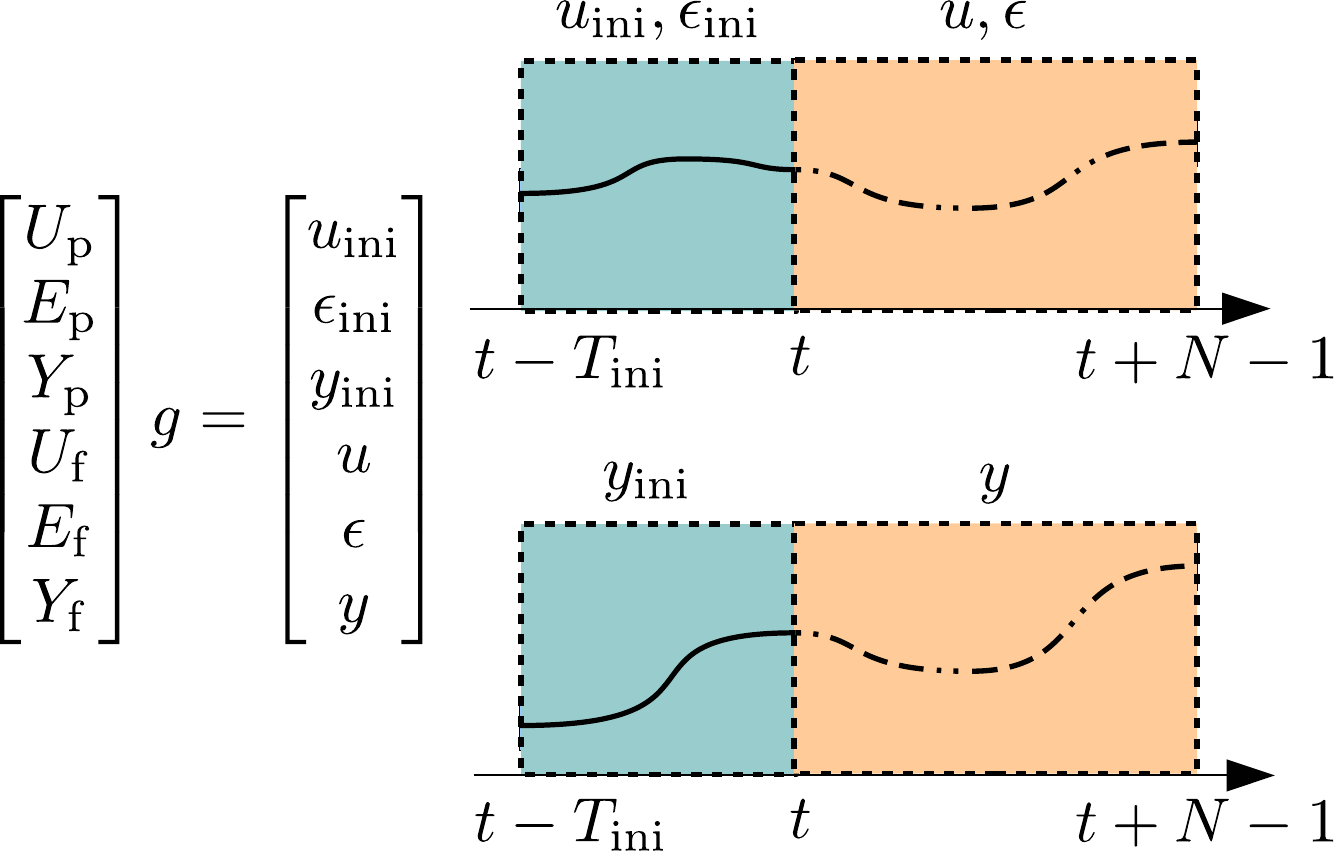}
	}
	\caption{Interpretation of Propositions 1 and 2. (a) A consecutive length-$T$ input/output trajectory is collected from the mixed traffic system. (b) All consecutive length-$(T_{\mathrm{ini}}+N)$ trajectories are extracted to construct the Hankel matrix (Proposition 1) or the Page matrix (Proposition 2). In particular, each trajectory is divided into two parts, i.e., ``previous data'' of length $T_{\mathrm{ini}}$ colored in cyan and ``future data'' of length $N$ colored in orange. (c) All valid length-$(T_{\mathrm{ini}}+N)$ trajectories of the mixed traffic system can be captured by the Hankel matrix or Page matrix.}
	\label{fig_proposition}
\end{figure*}	
	
Both Hankel matrix and Page matrix can be utilized to facilitate the non-parametric representation of dynamic systems. %The definition of these two matrix structures is provided now. 
Given a signal $\omega \in \mathbb{R}^{q}$ and two integers $i, j\in \mathbb{Z}_{+}$ with $i\le j$, we denote by $\omega_{\left[i, j\right]}$ the restriction of $\omega$ to the interval $\left[i, j\right]$, namely, $\omega_{\left[i, j\right]} := \begin{bmatrix}
	\omega^{\top}(i), \omega^{\top}(i+1), \cdots, \omega^{\top}(j)
\end{bmatrix}^{\top}$. To simplify notation, we will also use $\omega_{\left[i, j\right]}$ to denote the sequence $\left\lbrace \omega(i), \cdots, \omega(j)\right\rbrace$. The Hankel matrix of depth $k$ associated with $\omega_{\left[i, j\right]}$ is defined as
\[
\mathscr{H}_{k}(\omega_{\left[i, j\right]}):=\colvec{
	\omega(i) & \omega(i+1) & \cdots & \omega(j-k+1) \\
	\omega(i+1) & \omega(i+2) & \cdots & \omega(j-k+2)\\
	\vdots & \vdots & \ddots & \vdots \\
	\omega(i+k-1) & \omega(i+k) & \cdots & \omega(j)
}.
\]
Meanwhile, the Page matrix of depth $k$ associated with $\omega_{\left[i, j\right]}$ is defined as
\[
\begin{aligned}
&\mathscr{P}_{k}(\omega_{\left[i, j\right]}):=
\\
&\colvec{
	\omega(i) & \omega(i+k) & \cdots & \omega(i+\lfloor \frac{j-i+1}{k}-1 \rfloor k) \\
	\omega(i+1) & \omega(i+k+1) & \cdots & \omega(i+\lfloor \frac{j-i+1}{k}-1 \rfloor k+1)\\
	\vdots & \vdots & \ddots & \vdots \\
	\omega(i+k-1) & \omega(i+2k-1) & \cdots & \omega(i+\lfloor \frac{j-i+1}{k} \rfloor k-1)
},
\end{aligned}
\]
where $\lfloor \cdot \rfloor$ is the floor function which rounds its argument down to the nearest integer.

\begin{definition} [{\!\cite{Coulson2021TAC}}]
The sequence $\omega_{\left[i, j\right]}$ is said to be 
\begin{itemize}
	\item Hankel exciting of order $k$ if the matrix $\mathscr{H}_{k}(\omega_{\left[i, j\right]})$ has full row rank;
	\item $k$-Page exciting of order $l$, where $l\in \mathbb{Z}_{+}$, if the matrix
	\[
	\begin{bmatrix}
		\mathscr{P}_{k}(\omega_{\left[i, j-(l-1)k\right]})
		\\
		\mathscr{P}_{k}(\omega_{\left[k+i, j-(l-2)k\right]})
		\\
		\vdots
		\\
		\mathscr{P}_{k}(\omega_{\left[k(l-1)+i, j\right]})
	\end{bmatrix}
	\] 
	has full row rank.
\end{itemize}
\end{definition}
	
We now discuss the non-parametric representation of mixed traffic system \eqref{Eq:DT_TrafficModel}. It begins with data collection. Specifically, a trajectory of length $T \in \mathbb{Z}_{+}$ is collected from the mixed traffic system \eqref{Eq:DT_TrafficModel}, which includes the following two parts:
\begin{itemize}
	\item a combined input sequence of the mixed traffic system 
	\[
	\hat{u}^\mathrm{d}_{\left[1, T\right]}
		= \begin{bmatrix}
			\hat{u}^\mathrm{d}(1) \\ \vdots \\ \hat{u}^\mathrm{d}(T)
		\end{bmatrix} \in \mathbb{R}^{(m+1)T},
	\] which is comprised of CAVs' acceleration sequence and the velocity error sequence of the head vehicle, i.e.,  
	\[
	u^\mathrm{d}_{\left[1, T\right]}=\begin{bmatrix}
		u^\mathrm{d}(1) \\ \vdots \\ u^\mathrm{d}(T)
	\end{bmatrix} \in \mathbb{R}^{mT},  \epsilon^\mathrm{d}_{\left[1, T\right]}=\begin{bmatrix}
		\epsilon^\mathrm{d}(1) \\ \vdots \\ \epsilon^\mathrm{d}(T)
	\end{bmatrix} \in \mathbb{R}^{T};
	\] 
	\item the corresponding output sequence 
	\[y ^\mathrm{d}_{\left[1, T\right]}=\begin{bmatrix} 
		y^\mathrm{d}(1) \\ \vdots \\ y^\mathrm{d}(T) 
	\end{bmatrix} \in \mathbb{R}^{(n+m)T}.
	\]
\end{itemize}
After collecting these data sequences, different matrix structures can be used to store them for non-parametric representation. The Hankel matrix is the commonly used one~\cite{Wang2022arXiv,Coulson2019ECC}. Specifically, under the Hankel matrix structure, the data sequences are divided into two parts, i.e., the ``previous data'' of length $T_{\mathrm{ini}} \in \mathbb{Z}_{+}$ and ``future data'' of length $N \in \mathbb{Z}_{+}$, to construct the following matrices:
\begin{align}
	\begin{bmatrix}
		U_{\mathrm{p}}^{\mathscr{H}} \\
		U_{\mathrm{f}}^{\mathscr{H}}
	\end{bmatrix}&:=\mathscr{H}_{T_{\mathrm{ini}}+N}\left(u^{\mathrm{d}}_{\left[1, T\right]} \right), \begin{bmatrix}
		E_{\mathrm{p}}^{\mathscr{H}} \\
		E_{\mathrm{f}}^{\mathscr{H}}
	\end{bmatrix}:=\mathscr{H}_{T_{\mathrm{ini}}+N}\left(\epsilon^{\mathrm{d}}_{\left[1, T\right]} \right), \nonumber \\
	\begin{bmatrix}
		Y_{\mathrm{p}}^{\mathscr{H}} \\
		Y_{\mathrm{f}}^{\mathscr{H}}
	\end{bmatrix}&:=\mathscr{H}_{T_{\mathrm{ini}}+N}\left(y^{\mathrm{d}}_{\left[1, T\right]} \right), 	\label{Eq:DataHankel}
\end{align}
where $U_{\mathrm{p}}^{\mathscr{H}}$ denotes the first $T_{\mathrm{ini}}$ block rows of $\mathscr{H}_{T_{\mathrm{ini}}+N}(u^{\mathrm{d}}_{\left[1, T\right]})$ and $U_{\mathrm{f}}^{\mathscr{H}}$ denotes the last $N$ block rows of $\mathscr{H}_{T_{\mathrm{ini}}+N}(u^{\mathrm{d}}_{\left[1, T\right]})$, respectively (similarly for $E_{\mathrm{p}}^{\mathscr{H}}, E_{\mathrm{f}}^{\mathscr{H}}$ and $Y_{\mathrm{p}}^{\mathscr{H}}, Y_{\mathrm{f}}^{\mathscr{H}}$). 

Let $u_{\mathrm{ini}}=u_{\left[t-T_{\mathrm{ini}}, t-1\right]}$ be the control input sequence within a past time horizon of length $T_{\mathrm{ini}}$, and $u= u_{\left[t, t+N-1\right]}$ be the control input sequence within a prediction horizon of length $N$ (similarly for $\epsilon_\mathrm{ini},\epsilon$ and $y_\mathrm{ini},y$). Based on Willems' fundamental lemma~\cite{Willems2005SCL} and the \method{DeePC}~\cite{Coulson2019ECC}, we have the following proposition.
\begin{proposition} \label{Prop:Hankel}
    Consider a controllable mixed traffic system~\eqref{Eq:DT_TrafficModel} and assume that the combined input sequence $\hat{u}^\mathrm{d}_{\left[1, T\right]}$ is Hankel exciting of order $T_{\mathrm{ini}}+N+2n$. Then, any trajectory $(u_\mathrm{ini},\epsilon_\mathrm{ini},y_\mathrm{ini},u,\epsilon,y)$ of the mixed traffic system~\eqref{Eq:DT_TrafficModel} can be constructed via
	%Suppose the combined input sequence $\hat{u}^\mathrm{d}_{\left[1, T\right]}$ is Hankel exciting of order $T_{\mathrm{ini}}+N+2n$. Then a trajectory $(u_\mathrm{ini},\epsilon_\mathrm{ini},y_\mathrm{ini},u,\epsilon,y)$ belongs to the mixed traffic system ~\eqref{Eq:DT_TrafficModel} if there exists $g\in \mathbb{R}^{T-T_\mathrm{ini}-N+1}$ such that 
	%\vspace{-2mm}
	\begin{equation}
		\label{Eq:AdaptedDeePCAchievability}
		\begin{bmatrix}
			U_\mathrm{p}^{\mathscr{H}} \\ E_\mathrm{p}^{\mathscr{H}} \\ Y_\mathrm{p}^{\mathscr{H}} \\ U_\mathrm{f}^{\mathscr{H}} \\ E_\mathrm{f}^{\mathscr{H}} \\ Y_\mathrm{f}^{\mathscr{H}}
		\end{bmatrix}g=
		\begin{bmatrix}
			u_\mathrm{ini} \\ \epsilon_\mathrm{ini}\\ y_\mathrm{ini} \\ u \\\epsilon \\ y
		\end{bmatrix},
	\end{equation}
    where $g\in \mathbb{R}^{T-T_\mathrm{ini}-N+1}$.
\end{proposition}
		
Previous work \cite{Wang2022arXiv} relies on arranging the data sequences into Hankel matrices for non-parametric representation of the mixed traffic system. Here, we also explore another effective structure, i.e., the Page matrix structure, to arrange the data sequences. Similar to \eqref{Eq:DataHankel}, the data sequences are partitioned into two parts using Page matrices, as follows:
\begin{align}
	\begin{bmatrix}
		U_{\mathrm{p}}^{\mathscr{P}} \\
		U_{\mathrm{f}}^{\mathscr{P}}
	\end{bmatrix}&:=\mathscr{P}_{T_{\mathrm{ini}}+N}\left(u^{\mathrm{d}}_{\left[1, T\right]} \right), \begin{bmatrix}
		E_{\mathrm{p}}^{\mathscr{P}} \\
		E_{\mathrm{f}}^{\mathscr{P}}
	\end{bmatrix}:=\mathscr{P}_{T_{\mathrm{ini}}+N}\left(\epsilon^{\mathrm{d}}_{\left[1, T\right]} \right), \nonumber \\
	\begin{bmatrix}
		Y_{\mathrm{p}}^{\mathscr{P}} \\
		Y_{\mathrm{f}}^{\mathscr{P}}
	\end{bmatrix}&:=\mathscr{P}_{T_{\mathrm{ini}}+N}\left(y^{\mathrm{d}}_{\left[1, T\right]} \right), \label{Eq:DataPage}
\end{align}
where $U_{\mathrm{p}}^{\mathscr{P}}$ denotes the first $T_{\mathrm{ini}}$ block rows of $\mathscr{P}_{T_{\mathrm{ini}}+N}(u^{\mathrm{d}}_{\left[1, T\right]})$ and $U_{\mathrm{f}}^{\mathscr{P}}$ denotes the last $N$ block rows of $\mathscr{P}_{T_{\mathrm{ini}}+N}(u^{\mathrm{d}}_{\left[1, T\right]})$, respectively (similarly for $E_{\mathrm{p}}^{\mathscr{P}}, E_{\mathrm{f}}^{\mathscr{P}}$ and $Y_{\mathrm{p}}^{\mathscr{P}}, Y_{\mathrm{f}}^{\mathscr{P}}$). Based on Theorem 2.1 in \cite{Coulson2021TAC}, the non-parametric representation of the mixed traffic system can be formulated with the Page matrix structure as follows:
\begin{proposition} \label{Prop:Page}
    Consider a controllable mixed traffic system~\eqref{Eq:DT_TrafficModel} and assume that the combined input sequence $\hat{u}^\mathrm{d}_{\left[1, T\right]}$ is $(T_{\mathrm{ini}}+N)$-Page exciting of order $2n+1$. Then, any trajectory $(u_\mathrm{ini},\epsilon_\mathrm{ini},y_\mathrm{ini},u,\epsilon,y)$ of the mixed traffic system~\eqref{Eq:DT_TrafficModel} can be constructed via
	%Suppose the combined input sequence $\hat{u}^\mathrm{d}_{\left[1, T\right]}$ is $(T_{\mathrm{ini}}+N)$-Page exciting of order $2n+1$. Then a trajectory $(u_\mathrm{ini},\epsilon_\mathrm{ini},y_\mathrm{ini},u,\epsilon,y)$ belongs to the mixed traffic system ~\eqref{Eq:DT_TrafficModel} if there exists $g\in \mathbb{R}^{\lfloor \frac{T}{T_{\mathrm{ini}+N}} \rfloor}$ such that 
	%\vspace{-2mm}
	\begin{equation}
		\label{Eq:PageDeePCAchievability}
		\begin{bmatrix}
			U_\mathrm{p}^{\mathscr{P}} \\ E_\mathrm{p}^{\mathscr{P}} \\ Y_\mathrm{p}^{\mathscr{P}} \\ U_\mathrm{f}^{\mathscr{P}} \\ E_\mathrm{f}^{\mathscr{P}} \\ Y_\mathrm{f}^{\mathscr{P}}
		\end{bmatrix}g=
		\begin{bmatrix}
			u_\mathrm{ini} \\ \epsilon_\mathrm{ini}\\ y_\mathrm{ini} \\ u \\\epsilon \\ y
		\end{bmatrix},
	\end{equation}
    where $g\in \mathbb{R}^{\lfloor \frac{T}{T_{\mathrm{ini}+N}} \rfloor}$.
\end{proposition}
		
Fig.~\ref{fig_proposition} shows a schematic of Propositions~\ref{Prop:Hankel} and \ref{Prop:Page}. Propositions \ref{Prop:Hankel} and \ref{Prop:Page} reveal that by collecting sufficiently rich traffic data, one can directly predict the valid trajectories of the mixed traffic system without requiring an explicit model of HDVs.
 
\begin{remark}
	To satisfy the Hankel excitation condition in Proposition \ref{Prop:Hankel}, one should collect a minimum of $(m+2)(T_{\mathrm{ini}}+N+2n)-1$ data samples, i.e., $T\ge (m+2)(T_{\mathrm{ini}}+N+2n)-1$; on the other hand, to satisfy the Page excitation condition in Proposition \ref{Prop:Page}, one requires that $T\ge (T_{\mathrm{ini}}+N)(((m+1)(T_{\mathrm{ini}}+N)+1)(2n+1)-1)$.     \hfill $\square$
\end{remark}
		
\begin{remark} \label{remark_Page}
	The main difference between the Hankel matrix and the Page matrix is that none of the entries in the latter are repeated. This has both advantages and disadvantages. The main disadvantage is that more data samples are needed to construct the matrix, and thus the data collection is more expensive. However, more data samples usually contain more system information, which is conducive to achieving non-parametric system representation. In addition, if the measurements are corrupted by noise, the entries of the Page matrix are statistically independent, leading to algorithmically favourable properties, e.g., singular value decomposition can be used for de-noising~\cite{Huang2021TCST}.     \hfill $\square$
\end{remark}

\subsection{\method{DeeP-LCC} Formulation} \label{subsec_deepLcc}
	
Propositions \ref{Prop:Hankel} and \ref{Prop:Page} reveal that if sufficiently rich traffic data is collected, the future trajectory of the mixed traffic system can be directly predicted without relying on explicit system model. The relations \eqref{Eq:AdaptedDeePCAchievability} and \eqref{Eq:PageDeePCAchievability} are the non-parametric representations of the mixed traffic system, which are the key in formulating the \method{DeeP-LCC}. 

More precisely, at each time step $t$, the control actions of the CAVs are generated by solving the following optimization problem:
\vspace{-3pt}
\begin{equation} \label{Eq:AdaptedDeePC}
	\begin{aligned}
		\min_{g,u,y} \quad & J(y,u) = \sum\limits_{k=t}^{t+N-1}\left( \left\|y(k)\right\|_{Q}^{2}+\left\|u(k)\right\|_{R}^{2}\right)
		\\
		\mathrm{s.t.} \quad & \eqref{Eq:AdaptedDeePCAchievability} \; (\mathrm{or} \; \eqref{Eq:PageDeePCAchievability}),
		\\
		& \epsilon = 0_N,
		\\
		&y_\mathrm{min} \leq y(k) \leq y_\mathrm{max}, \quad k=t, \ldots, t+N-1,
		\\
		&a_\mathrm{min} \leq u(k) \leq a_\mathrm{max}, \quad k=t, \ldots, t+N-1.
	\end{aligned}
\end{equation}
In \eqref{Eq:AdaptedDeePC}, $J(y,u)$ is a quadratic cost function, and its weighting matrices $Q$ and $R$ are selected as $Q=\mathrm{diag}(Q_C, Q_{H}) \in \mathbb{R}^{(n+m)\times (n+m)}$ with $Q_{C}=\mathrm{diag}(w_{s},w_{v},\ldots,w_{s},w_{v})\in \mathbb{R}^{2m\times 2m}$, $Q_{H}=\mathrm{diag}(w_{v},\ldots,w_{v}) \in \mathbb{R}^{(n-m)\times (n-m)}$ and $R=\mathrm{diag}(w_{u},\ldots,w_{u})\in \mathbb{R}^{m\times m}$, where $w_{s}$, $w_{v}$, $w_{u}$ are the weighting factors for the spacing error of CAVs, the velocity error of all vehicles, and the control input of CAVs, respectively. The second constraint,  $\epsilon = 0_N$, is used to predict the future velocity error sequence of the head vehicle. The third constraint is applied to the output of the mixed traffic system for safety guarantees, and the lower and upper bounds of the output signal are defined as 
$$
\begin{aligned}
y_{\mathrm{min}} &=\begin{bmatrix}
(1_{m}\otimes \begin{bmatrix}
\tilde{s}_{\mathrm{min}}, \tilde{v}_{\mathrm{min}}
\end{bmatrix}^{\top})^{\top}, (1_{n-m}\otimes \tilde{v}_{\mathrm{min}})^{\top}
\end{bmatrix}^{\top} \\
y_{\mathrm{max}}&=\begin{bmatrix}
(1_{m}\otimes \begin{bmatrix}
	\tilde{s}_{\mathrm{max}}, \tilde{v}_{\mathrm{max}}
\end{bmatrix}^{\top})^{\top}, (1_{n-m}\otimes \tilde{v}_{\mathrm{max}})^{\top}
\end{bmatrix}^{\top}
\end{aligned}
$$ with $\tilde{s}_{\mathrm{min}}$, $\tilde{s}_{\mathrm{max}}$ ($\tilde{v}_{\mathrm{min}}$, $\tilde{v}_{\mathrm{max}}$) being the lower and upper bounds of spacing (velocity) error. The fourth constraint in \eqref{Eq:AdaptedDeePC} is applied to the input of the mixed traffic system, and $a_{\mathrm{min}}$, $a_{\mathrm{max}}$ denote the lower and upper bounds of input signal. We refer the interested reader to \cite{Wang2022arXiv} for more details on designing the cost function and constraints in \method{DeeP-LCC}. 
		
\section{Privacy-Preserving \method{DeeP-LCC}} \label{sec_privacyPreserved}
		
In this section, we first introduce the attack models and then present an affine masking strategy. We finally propose a new privacy-preserving \method{DeeP-LCC}. 
		
\subsection{Attack Models} \label{subsec_attackModel}
\begin{figure}[t]
	\centering
	\includegraphics[width=0.43\textwidth]{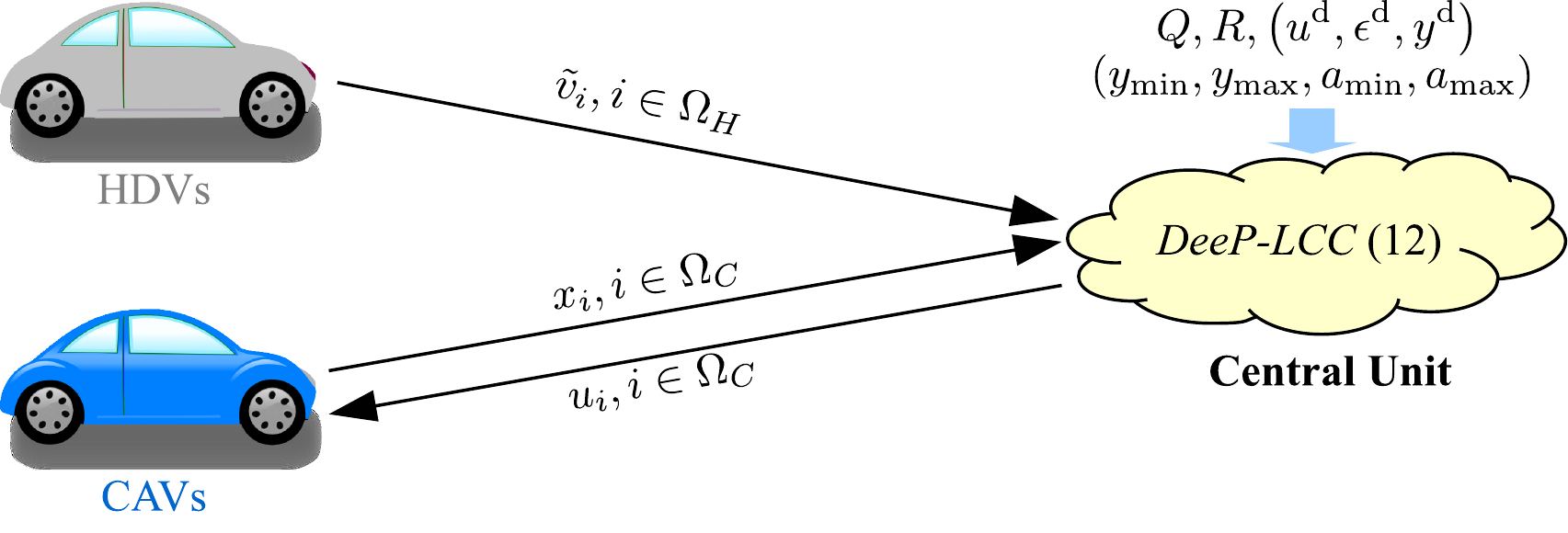}
	\caption{Unsecure \method{DeeP-LCC} architecture.}
	\label{Fig:architecture}
\end{figure}

For the mixed-traffic vehicle fleet discussed above, a central unit (e.g., a road-side edge compute or a remote cloud) is needed to receive all the vehicle data and solve the optimization problem \eqref{Eq:AdaptedDeePC}. A feasible architecture is shown in Fig.~\ref{Fig:architecture} and is described as follows:
\begin{itemize}
	\item \textit{Handshaking Phase:} The vehicle system sends 
 $$\left\lbrace Q, R, (u^{\mathrm{d}}_{\left[1, T\right]}, \epsilon^{\mathrm{d}}_{\left[1, T\right]}, y^{\mathrm{d}}_{\left[1, T\right]}), (y_{\mathrm{min}}, y_{\mathrm{max}}, a_{\mathrm{min}}, a_{\mathrm{max}}) \right\rbrace$$ to the central unit, which are the necessary information for the central unit to set up the optimization problem~\eqref{Eq:AdaptedDeePC}.
			
	\item \textit{Execution Phase:} At each time step $k$, each CAV and HDV sends its state $x_{i}(k)=\begin{bmatrix}
	\tilde{s}_{i}(k), \tilde{v}_{i}(k)
	\end{bmatrix}^{\top}$ ($i\in \Omega_{C}$) and velocity error $\tilde{v}_{i}(k)$ ($i\in \Omega_{H}$), respectively, to the central unit. The central unit computes $u(k)$ by solving the optimization problem \eqref{Eq:AdaptedDeePC} and sends optimal $u_{i}(k)$ ($i\in \Omega_{C}$) to the CAVs. Finally, the CAVs applies $u_{i}(k)$ to the actuators and the system evolves over one step.
\end{itemize}
		
The involved vehicles need to provide the central unit with real-time measurements, pre-collected data, cost function, and constraints to facilitate the calculation of \eqref{Eq:AdaptedDeePC}. The data may contain private contents that need to be protected from external attackers. In this paper, we consider two attack models \cite{Gao2022TITS}:
\begin{itemize}
	\item \emph{Eavesdropping attacks} are attacks in which an external eavesdropper wiretaps communication channels to intercept exchanged messages in an attempt to learn the information about sending parties.
	\item \emph{Honest-but-curious attacks} are attacks in which the untrusted central unit follows the protocol steps correctly but is curious and collects received intermediate data in an attempt to learn the information about the vehicles.
\end{itemize}
		
In particular, we consider the case that the privacy-sensitive information is contained in the system state and input of CAVs, i.e., $x_{i}(k)=\begin{bmatrix}
	\tilde{s}_{i}(k), \tilde{v}_{i}(k)
\end{bmatrix}^{\top}, u_{i}(k), i\in \Omega_{C}$. It is apparent that the attacker can successfully eavesdrop the messages $x_{i}(k)$ and $u_{i}(k)$ when the \method{DeeP-LCC} architecture introduced in Section \ref{subsec_deepLcc} is adopted. 
		
In the next subsection, we develop a masking mechanism to modify the exchanged information between the CAVs and the central unit such that an equivalent data-enabled predictive control problem is solved without affecting system performance while preventing the attacker from wiretapping the CAVs' state and input. Although we only consider the privacy preservation for CAVs, the following proposed approach can also be straightforwardly adopted by HDVs to mask privacy-sensitive information, e.g., velocity error $\tilde{v}_{i}(k)$, $i\in \Omega_{H}$.  
		
\subsection{Affine Masking} \label{subsec_affine}
		
Affine masking is a class of algebraic transformations and recently has been adopted to accomplish privacy protection for cloud-based control \cite{Sultangazin2020TAC,Zhang2021privacy}. In this paper, by considering the characteristics of the mixed traffic system, we introduce affine transformation maps to design a privacy-preserved \method{DeeP-LCC} scheme. Specifically, for each CAV, two invertible affine maps are employed to transform the state $x_{i}(k)$ and input $u_{i}(k)$ to the new state $\bar{x}_{i}(k)$ and input $\bar{u}_{i}(k)$, as follows:
\begin{equation} \label{Eq:Affine}
	\begin{aligned}
		\begin{cases}
			\bar{x}_{i}(k) =  P_{x,i}x_{i}(k)+l_{x,i},
			\\
			\bar{u}_{i}(k) = P_{u,i}u_{i}(k)+l_{u,i},
		\end{cases} \quad \forall i\in \Omega_{C},
	\end{aligned}
\end{equation}
where $P_{x,i}\in \mathbb{R}^{2\times 2}$ is an arbitrary invertible matrix, and $l_{x,i}\in \mathbb{R}^{2}$, $P_{u,i}\in \mathbb{R}$ and $l_{u,i}\in \mathbb{R}$ are arbitrary non-zero vector or constant with compatible dimensions. After each CAV masks its true state and input via \eqref{Eq:Affine}, the new input and output signals of the mixed traffic system are defined as
\begin{align}
	\bar{u}(k)&=\begin{bmatrix}
		\bar{u}_{i_1}(k), \bar{u}_{i_2}(k), \cdots, \bar{u}_{i_m}(k)
	\end{bmatrix}^{\top}, \label{Eq:SystemOutputAffine} \\
	\bar{y}(k)&=\begin{bmatrix}
		\bar{x}_{i_{1}}^{\top}(k),\ldots, \bar{x}_{i_{m}}^{\top}(k),
		\tilde{v}_{j_{1}}(k),\ldots,\tilde{v}_{j_{n-m}}(k)
	\end{bmatrix}^{\top}. \nonumber
\end{align}

Based on \eqref{Eq:Affine}, \eqref{Eq:SystemOutputAffine} and the definition of $u(k)$ and $y(k)$, we obtain that
\begin{equation} \label{Eq:uy_bar}
	\bar{u}(k) = P_{u}u(k) + L_{u}, \quad \bar{y}(k) = P_{y}y(k) + L_{y},
\end{equation}
where 
\[
\begin{aligned}
	P_{u} &= \mathrm{diag}(P_{u,i_{1}}, P_{u,i_{2}}, \ldots, P_{u,i_{m}})\in \mathbb{R}^{m\times m}, 
	\\
	L_{u} &= \begin{bmatrix}
		l_{u,i_{1}}, l_{u,i_{2}}, \ldots, l_{u,i_{m}}
	\end{bmatrix}^{\top}\in \mathbb{R}^{m},
	\\
	P_{y} &= \mathrm{diag}(P_{x,i_{1}}, \ldots, P_{x,i_{m}}, I_{n-m})\in \mathbb{R}^{(n+m)\times (n+m)},
	\\
	L_{y} &= \begin{bmatrix}
		l_{x,i_{1}}^{\top}, \ldots, l_{x,i_{m}}^{\top}, 0_{n-m}^{\top}
	\end{bmatrix}^{\top} \in \mathbb{R}^{(n+m)}.
\end{aligned}
\]
In \eqref{Eq:uy_bar}, $(P_{u}, L_{u})$ and $(P_{y}, L_{y})$ are two affine maps constructed from the local affine maps of CAVs, and are used to transform the original input and output of mixed traffic system, i.e., $(u(k), y(k))$, into the new ones $(\bar{u}(k), \bar{y}(k))$. Given the affine transformation, the discrete-time model of the mixed traffic system \eqref{Eq:DT_TrafficModel} can be reformulated as follows:
\begin{equation} \label{Eq:DT_TrafficModelAffine}
	\begin{cases}
		x(k+1) = A_\mathrm{d}x(k) + \bar{B}_\mathrm{d}\bar{u}(k) + H_\mathrm{d} \epsilon(k) + \bar{L}_{u},\\
		\bar{y}(k) = \bar{C}_\mathrm{d}x(k) + L_{y},
	\end{cases}
\end{equation}
where $\bar{B}_{\mathrm{d}}=B_{\mathrm{d}}P_{u}^{-1}$, $\bar{L}_{u}=-B_{\mathrm{d}}P_{u}^{-1}L_{u}$, and $\bar{C}_{\mathrm{d}}=P_{y}C_{\mathrm{d}}$.
		
Under the affine transformation mechanism \eqref{Eq:SystemOutputAffine}, the exchanged information between the CAVs and the central unit changes to $\bar{x}_{i}(k), \bar{u}_{i}(k), i\in \Omega_{C}$, which can avoid leaking the real state and input signals to the eavesdropper or the central unit. This affine transformation mechanism also leads to a new system formulation \eqref{Eq:DT_TrafficModelAffine}, and thus a compatible \method{DeeP-LCC} method needs to be developed.
		
\subsection{Privacy-Preserving \method{DeeP-LCC} Reformulation} \label{subsec_reformulation}
		
Denote $\bar{u}^{\mathrm{d}}_{\left[1, T\right]}=\begin{bmatrix}
	\bar{u}^{\mathrm{d}}(1),\ldots,\bar{u}^{\mathrm{d}}(T)
\end{bmatrix}^{\top}$ as the corresponding acceleration sequence of $u^{\mathrm{d}}_{\left[1, T\right]}$ under the affine map $(P_{u}, L_{u})$, and $\bar{y}^{\mathrm{d}}_{\left[1, T\right]} =\begin{bmatrix}
	\bar{y}^\mathrm{d}(1),\ldots,\bar{y}^\mathrm{d}(T)
\end{bmatrix}^{\top}$ as the corresponding output sequence of $y^{\mathrm{d}}_{\left[1, T\right]}$ under the affine map $(P_{y}, L_{y})$. Then, similar to $\hat{u}^\mathrm{d}_{\left[1, T\right]}$, the combined input sequence $\hat{\bar{u}}^\mathrm{d}_{\left[1, T\right]}\in \mathbb{R}^{(m+1)T}$ is constructed with $\bar{u}^{\mathrm{d}}_{\left[1, T\right]}$ and $\epsilon^{\mathrm{d}}_{\left[1, T\right]}$. The matrices $(\bar{U}_{\mathrm{p}}^{\mathscr{H}}, \bar{U}_{\mathrm{f}}^{\mathscr{H}}, \bar{Y}_{\mathrm{p}}^{\mathscr{H}}, \bar{Y}_{\mathrm{f}}^{\mathscr{H}})$ and $(\bar{U}_{\mathrm{p}}^{\mathscr{P}}, \bar{U}_{\mathrm{f}}^{\mathscr{P}}, \bar{Y}_{\mathrm{p}}^{\mathscr{P}}, \bar{Y}_{\mathrm{f}}^{\mathscr{P}})$ can be constructed with $\bar{u}^{\mathrm{d}}_{\left[1, T\right]}$ and $\bar{y}^{\mathrm{d}}_{\left[1, T\right]}$ by following the same procedure shown in \eqref{Eq:DataHankel} and \eqref{Eq:DataPage}, respectively. Motivated by Willems' fundamental lemma~\cite{Willems2005SCL}, \eqref{Eq:AdaptedDeePCAchievability} and \eqref{Eq:PageDeePCAchievability}, we use the data $\left( \bar{u}^{\mathrm{d}}_{\left[1, T\right]}, \epsilon^{\mathrm{d}}_{\left[1, T\right]}, \bar{y}^{\mathrm{d}}_{\left[1, T\right]} \right)$ to represent the affine masking-based mixed traffic system \eqref{Eq:DT_TrafficModelAffine} under the Hankel and Page matrix structures. The results are summarized below.
		
\begin{proposition} \label{Prop:Hankel_privacy}
	Suppose the data $\hat{\bar{u}}^\mathrm{d}_{\left[1, T\right]}$ is Hankel exciting of order $T_{\mathrm{ini}}+N+2n+1$. Then, any trajectory $ (\bar{u}_\mathrm{ini},\epsilon_\mathrm{ini},\bar{y}_\mathrm{ini},\bar{u},\epsilon,\bar{y})$ of \eqref{Eq:DT_TrafficModelAffine} can be constructed via
	\begin{equation}
		\label{Eq:AdaptedDeePCAchievabilityAffine_privacy}
		\begin{bmatrix}
			\bar{U}_\mathrm{p}^{\mathscr{H}} \\ E_\mathrm{p}^{\mathscr{H}} \\ \bar{Y}_\mathrm{p}^{\mathscr{H}} \\ \bar{U}_\mathrm{f}^{\mathscr{H}} \\ E_\mathrm{f}^{\mathscr{H}} \\ \bar{Y}_\mathrm{f}^{\mathscr{H}} \\ 1_{T-T_{\mathrm{ini}}-N+1}^{\top}
		\end{bmatrix}\bar{g}=
		\begin{bmatrix}
			\bar{u}_\mathrm{ini} \\ \epsilon_\mathrm{ini}\\ \bar{y}_\mathrm{ini} \\ \bar{u} \\\epsilon \\ \bar{y} \\ 1
		\end{bmatrix},
	\end{equation}
    where $\bar{g}\in \mathbb{R}^{T-T_\mathrm{ini}-N+1}$.
\end{proposition}
		
\begin{IEEEproof}
	See Appendix \ref{Appen:Prof_Hankel}.
\end{IEEEproof}
		
\begin{proposition} \label{Prop:Page_privacy}
	Suppose the data $\hat{\bar{u}}^\mathrm{d}_{\left[1, T\right]}$ is $(T_{\mathrm{ini}}+N)$-Page exciting of order $2n+2$. Then, any trajectory $ (\bar{u}_\mathrm{ini},\epsilon_\mathrm{ini},\bar{y}_\mathrm{ini},\bar{u},\epsilon,\bar{y})$ of \eqref{Eq:DT_TrafficModelAffine} can be constructed via 
	\begin{equation}
		\label{Eq:PageDeePCAchievabilityAffine_privacy}
		\begin{bmatrix}
			\bar{U}_\mathrm{p}^{\mathscr{P}} \\ E_\mathrm{p}^{\mathscr{P}} \\ \bar{Y}_\mathrm{p}^{\mathscr{P}} \\ \bar{U}_\mathrm{f}^{\mathscr{P}} \\ E_\mathrm{f}^{\mathscr{P}} \\ \bar{Y}_\mathrm{f}^{\mathscr{P}} \\ 1_{\lfloor \frac{T}{T_{\mathrm{ini}+N}} \rfloor}^{\top}
		\end{bmatrix}\bar{g}=
		\begin{bmatrix}
			\bar{u}_\mathrm{ini} \\ \epsilon_\mathrm{ini}\\ \bar{y}_\mathrm{ini} \\ \bar{u} \\\epsilon \\ \bar{y} \\ 1
		\end{bmatrix},
	\end{equation}
    where $\bar{g}\in \mathbb{R}^{\lfloor \frac{T}{T_{\mathrm{ini}+N}} \rfloor}$.
\end{proposition}
		
\begin{IEEEproof}
	See Appendix \ref{Appen:Prof_Page}.
\end{IEEEproof}
		
\begin{remark}
	Propositions \ref{Prop:Hankel} and \ref{Prop:Page} work for standard LTI system \eqref{Eq:DT_TrafficModel}, while
	Propositions \ref{Prop:Hankel_privacy} and \ref{Prop:Page_privacy} are extended to LTI system \eqref{Eq:DT_TrafficModelAffine} with non-zero offsets $\bar{L}_{u}$ and $L_{y}$. One key difference between Propositions \ref{Prop:Hankel}/\ref{Prop:Page} and \ref{Prop:Hankel_privacy}/\ref{Prop:Page_privacy} is the persistent excitation condition on $\hat{\bar{u}}^\mathrm{d}_{\left[1, T\right]}$. Proposition \ref{Prop:Hankel}/\ref{Prop:Page} requires $\hat{u}^\mathrm{d}_{\left[1, T\right]}$ to be Hankel exciting of order $T_{\mathrm{ini}}+N+2n$ ($(T_{\mathrm{ini}}+N)$-Page exciting of order $2n+1$), while Proposition \ref{Prop:Hankel_privacy}/\ref{Prop:Page_privacy} imposes that $\hat{u}^\mathrm{d}_{\left[1, T\right]}$ is Hankel exciting of order $T_{\mathrm{ini}}+N+2n+1$ ($(T_{\mathrm{ini}}+N)$-Page exciting of order $2n+2$). 
    The latter introduces an additional order to handle the constraint $1_{T-T_{\mathrm{ini}}-N+1}^{\top}\bar{g}=1$ ($1_{\lfloor \frac{T}{T_{\mathrm{ini}+N}} \rfloor}^{\top}\bar{g}=1$). This constaint ensures that the offsets $\bar{L}_{u}$ and $L_{y}$ will be carried through from the pre-collected data $\left( \bar{u}^{\mathrm{d}}, \epsilon^{\mathrm{d}}, \bar{y}^{\mathrm{d}} \right)$ to the trajectory $ (\bar{u}_\mathrm{ini},\epsilon_\mathrm{ini},\bar{y}_\mathrm{ini},\bar{u},\epsilon,\bar{y})$. Similar constraints can be found in \cite{Berberich2021arXiv,Salvador2019ECC}.  \hfill $\square$
	%The key difference between Proposition \ref{Prop:Hankel}(\ref{Prop:Page}) and \ref{Prop:Hankel_privacy} (\ref{Prop:Page_privacy}) is the condition $$1_{T-T_{\mathrm{ini}}-N+1}^{\top}\bar{g}=1, \qquad 1_{\lfloor \frac{T}{T_{\mathrm{ini}+N}} \rfloor}^{\top}\bar{g}=1.$$ This condition ensures that the offsets $\bar{L}_{u}$ and $L_{y}$ will be carried through from the pre-collected data $\left( \bar{u}^{\mathrm{d}}, \epsilon^{\mathrm{d}}, \bar{y}^{\mathrm{d}} \right)$ to the trajectory $ (\bar{u}_\mathrm{ini},\epsilon_\mathrm{ini},\bar{y}_\mathrm{ini},\bar{u},\epsilon,\bar{y})$. Similar conditions can be found in \cite{Berberich2021arXiv,Salvador2019ECC}.   \hfill $\square$ 
\end{remark}
		
The affine maps are able to mask the true system state $x_{i}(k)$ and input $u_{i}(k)$ of CAVs to protect the privacy, and in the central unit, a new optimization problem with respect to $\left(\bar{y},\bar{u}\right)$ and the new non-parametric representation \eqref{Eq:AdaptedDeePCAchievabilityAffine_privacy} (or \eqref{Eq:PageDeePCAchievabilityAffine_privacy}) are solved. Specifically, with the affine maps, one can show that \eqref{Eq:AdaptedDeePC} can be transformed into the following problem:
\begin{equation} \label{Eq:AdaptedDeePCAffine}
	\begin{aligned}
		\min_{\bar{g},\bar{u},\bar{y}} \quad & \bar{J}(\bar{y},\bar{u}) 
		\\
		%	&= \sum\limits_{k=t}^{t+N-1}\left( \left\|\bar{y}(k)\right\|_{\bar{Q}}^{2} + \bar{q}^{\top}\bar{y}(k) +\left\|\bar{u}(k)\right\|_{\bar{R}}^{2} + \bar{r}^{\top}\bar{u}(k) \right)
		%	\\
		\mathrm{s.t.} \quad &\eqref{Eq:AdaptedDeePCAchievabilityAffine_privacy} \; (\mathrm{or} \; \eqref{Eq:PageDeePCAchievabilityAffine_privacy}),
		\\
		& \epsilon = 0_N,
		\\
		&\bar{y}_\mathrm{min} \leq \bar{y}(k) \leq \bar{y}_\mathrm{max}, \quad k=t, \ldots, t+N-1,
		\\
		&\bar{a}_\mathrm{min} \leq \bar{u}(k) \leq \bar{a}_\mathrm{max}, \quad k=t, \ldots, t+N-1,
	\end{aligned}
\end{equation}
where the cost function is defined as 
$$\bar{J}(\bar{y},\bar{u}) \!= \!\!\sum\limits_{k=t}^{t+N-1}\left( \left\|\bar{y}(k)\right\|_{\bar{Q}}^{2} + \bar{q}^{\top}\bar{y}(k) +\left\|\bar{u}(k)\right\|_{\bar{R}}^{2} + \bar{r}^{\top}\bar{u}(k) \right) $$
with  $\bar{Q} \in \mathbb{R}^{(n+m)\times (n+m)}$, $\bar{q}\in \mathbb{R}^{n}$, $\bar{R}\in \mathbb{R}^{m\times m}$, $\bar{r}\in \mathbb{R}^{m}$, $\bar{Q}_{f}\in \mathbb{R}^{n\times n}$, and $\bar{q}_{f}\in \mathbb{R}^{n}$  given by
\begin{align}
	\bar{Q} &= P_{y}^{-\top}QP_{y}^{-1}\in \mathbb{R}^{(n+m)\times (n+m)}, 
	%\\
	\bar{q} = -2\bar{Q}L_{y} \in \mathbb{R}^{(n+m)}, \nonumber
	\\
	\bar{R} &= P_{u}^{-\top}RP_{u}^{-1}\in \mathbb{R}^{m\times m}, 
	%\\
	\bar{r} = -2\bar{R}L_{u}\in \mathbb{R}^{m}. \label{Eq:paraAffine}
\end{align}
In \eqref{Eq:AdaptedDeePCAffine}, $\left(\bar{y}_{\mathrm{min}}, \bar{y}_{\mathrm{max}}\right)$ and $\left(\bar{u}_{\mathrm{min}}, \bar{u}_{\mathrm{max}}\right)$ are the corresponding constraint bounds of $\left(y_{\mathrm{min}}, y_{\mathrm{max}}\right)$ and $\left(u_{\mathrm{min}}, u_{\mathrm{max}}\right)$ under the affine maps $\left( P_{y}, L_{y} \right)$ and $\left( P_{u}, L_{u} \right)$ given in \eqref{Eq:uy_bar}, i.e., $\bar{y}_{\mathrm{min}}=P_{y}y_{\mathrm{min}}+L_{y}$, $\bar{y}_{\mathrm{max}}=P_{y}y_{\mathrm{max}}+L_{y}$ and $\bar{u}_{\mathrm{min}}=P_{u}u_{\mathrm{min}}+L_{u}$, $\bar{u}_{\mathrm{max}}=P_{u}u_{\mathrm{max}}+L_{u}$.
		
\begin{figure}[t]
	\centering
	\includegraphics[width=0.43\textwidth]{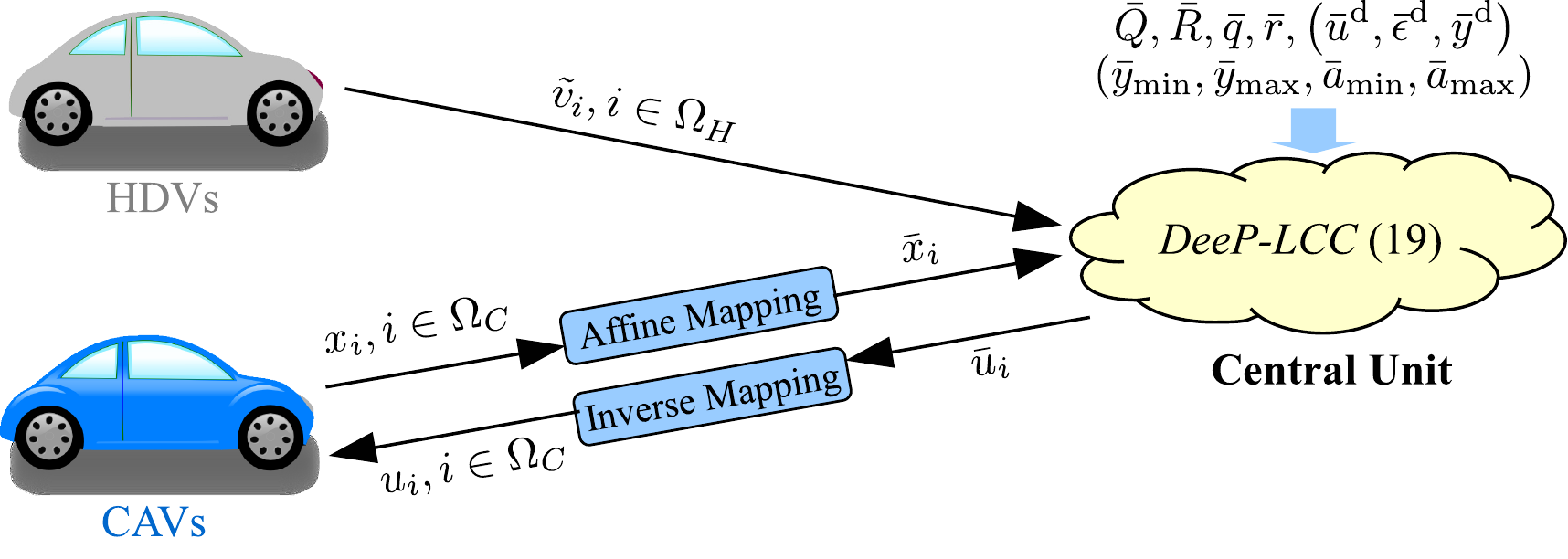}
	\caption{Privacy-preservation \method{DeeP-LCC} architecture.}
	\label{Fig:architecturePrivacy}
\end{figure}

Compared to the unsecure \method{DeeP-LCC} in Section~\ref{subsec_attackModel}, our privacy-preserved \method{DeeP-LCC} architecture with the affine maps, shown in Fig.~\ref{Fig:architecturePrivacy}, is modified as:
\begin{itemize}
	\item \textit{Handshaking Phase:} The vehicle system sends 
	\[
	\left\lbrace \! \bar{Q}, \bar{R}, \bar{q}, \bar{r}, (\bar{u}^{\mathrm{d}}_{\left[1, T\right]}, \epsilon^{\mathrm{d}}_{\left[1, T\right]}, \bar{y}^{\mathrm{d}}_{\left[1, T\right]}), (\bar{y}_{\mathrm{min}}, \bar{y}_{\mathrm{max}}, \bar{a}_{\mathrm{min}}, \bar{a}_{\mathrm{max}}) \!\right\rbrace
	\] 
	to the central unit, that is, the necessary information to set up the optimization problem \eqref{Eq:AdaptedDeePCAffine}.
			
	\item \textit{Execution Phase:} At each time step $k$, the CAVs encodes $x_{i}(k)$ ($i\in \Omega_{C}$) into $\bar{x}_{i}(k)$ with $\left( P_{x,i}, l_{x,i} \right)$ and sends $\bar{x}_{i}(k)$ to the central unit. Meanwhile, HDVs sends velocity error $\tilde{v}_{i}(k)$ ($i\in \Omega_{H}$) to the central unit. After receiving these data, the central unit computes $\bar{u}(k)$ by solving the optimization problem \eqref{Eq:AdaptedDeePCAffine} and sends the solution $\bar{u}_{i}(k)$ ($i\in \Omega_{C}$) to the CAVs. Finally, each CAV uses $\left( P_{u,i}, l_{u,i} \right)$ to decode $\bar{u}_{i}(k)$, i.e., $u_{i}(k) = P_{u,i}^{-1}\left( \bar{u}_{i}(k)-l_{u,i} \right)$ and applies $u_{i}(k)$ to the actuators. The system then evolves over one step.
\end{itemize}

Note that during the handshaking phase, the information received by the central unit does not include affine maps $\left( P_{y}, L_{y} \right)$ and $\left( P_{u}, L_{u} \right)$, and each CAV will not share its local affine map $\left( P_{u,i}, l_{u,i} \right)$ with others during the execution phase. Therefore, the attackers are unable to infer the privacy-sensitive information $x_{i}(k)$ and $u_{i}(k)$. Rigorous definition and analysis of privacy preservation will be provided in Section~\ref{subsec_privacy}. 
		
\begin{remark}
	The formulations \eqref{Eq:AdaptedDeePCAchievabilityAffine_privacy} and \eqref{Eq:PageDeePCAchievabilityAffine_privacy} are valid for deterministic LTI system \eqref{Eq:DT_TrafficModelAffine}. However, in practice, the car-following behavior of HDVs is stochastic and has uncertainties, which leads to a non-deterministic and nonlinear mixed traffic system. Inspired by the regularization design for standard \method{DeePC} \cite{Coulson2019ECC}, slack variable $\bar{\sigma}_{y} \in \mathbb{R}^{(n+m)T_{\mathrm{ini}}}$ and two-norm regularization on $\bar{g}$ can be introduced to handle system uncertainties and nonlinearities. For instance, the optimization problem \eqref{Eq:AdaptedDeePCAffine} under the Hankel matrix structure becomes
	\begin{equation} \label{Eq:DeePCAffineFea}
		\begin{aligned}
			\min_{\bar{g},\bar{u},\bar{y}} \quad & \bar{J}(\bar{y},\bar{u}) + \bar{\lambda}_{g} \left\|\bar{g}\right\|_{2}^{2} + \bar{\lambda}_{\sigma} \left\|\bar{\sigma}_{y}\right\|_{2}^{2}
			\\
			\mathrm{s.t.} \quad &\begin{bmatrix}
				\bar{U}_\mathrm{p} \\ E_\mathrm{p}\\ \bar{Y}_\mathrm{p} \\ \bar{U}_\mathrm{f} \\ E_\mathrm{f}\\ \bar{Y}_\mathrm{f} \\ 1_{T-T_{\mathrm{ini}}-N+1}
			\end{bmatrix}\bar{g}=
			\begin{bmatrix}
				\bar{u}_\mathrm{ini} \\ \epsilon_\mathrm{ini}\\ \bar{y}_\mathrm{ini} \\ \bar{u} \\\epsilon \\ \bar{y} \\ 1
			\end{bmatrix} + \begin{bmatrix}
				0 \\ 0\\ \bar{\sigma}_{y} \\ 0 \\ 0 \\ 0 \\ 0
			\end{bmatrix},
			\\
			& \epsilon = 0_N,
			\\
			&\bar{y}_\mathrm{min} \leq \bar{y}(k) \leq \bar{y}_\mathrm{max}, \quad k=t, \ldots, t+N-1,
			\\
			&\bar{a}_\mathrm{min} \leq \bar{u}(k) \leq \bar{a}_\mathrm{max}, \quad k=t, \ldots, t+N-1,
		\end{aligned}
	\end{equation}
	where $\bar{\lambda}_{g}>0$ and $\bar{\lambda}_{\sigma} >0$ are weighting coefficients. The slack variable $\bar{\sigma}_{y}$ for the past output signal is added to ensure the feasibility of the equality constraint, while the regularization on $\bar{g}$ is used to avoid overfitting. \hfill $\square$
\end{remark}
		
\section{Equivalence and Privacy Preservation} \label{sec_privacyAnalysis}

As mentioned in Section \ref{subsec_attackModel}, the attacker aims to infer the system state and input of CAVs, i.e., $x_{i}(k)=\begin{bmatrix}
	\tilde{s}_{i}(k), \tilde{v}_{i}(k)
\end{bmatrix}^{\top}, u_{i}(k), i\in \Omega_{C}$. Note that under the privacy-preserving \method{DeeP-LCC} architecture, the exchanged information between CAVs and the central unit during the execution phase is $\bar{x}_{i}(k)$ and $\bar{u}_{i}(k)$ rather than the actual system state and input.  
In the section, we first prove that the reformulated \method{DeeP-LCC} problem \eqref{Eq:AdaptedDeePCAffine} is equivalent to the original \method{DeeP-LCC} problem \eqref{Eq:AdaptedDeePC}, and then we show that the privacy of CAVs' state and input is indeed protected.
		
\subsection{Equivalence with Affine Transformation}
		
The following Lemma establishes that the equivalence of the reformulated \method{DeeP-LCC} problem \eqref{Eq:AdaptedDeePCAffine} to the original \method{DeeP-LCC} problem \eqref{Eq:AdaptedDeePC}. 
		
\begin{lemma} \label{Lemma}
	Under the affine transformation mechanism, the optimization problem \eqref{Eq:AdaptedDeePCAffine} is equivalent to \eqref{Eq:AdaptedDeePC}.
	%i.e., if {\small$\bar{\bm{U}}^{*}_{k}=\begin{bmatrix} \bar{u}_{0|k}^{*\top}, \cdots, \bar{u}_{N-1|k}^{*\top} \end{bmatrix}^{\top}$} is a local (resp. global) minimizer of $\textbf{P-2}$, then the transformed control via inverse mapping {\small$\bm{U}^{*}_{k} = \begin{bmatrix} u_{0|k}^{*\top}, \cdots, u_{N-1|k}^{*\top} \end{bmatrix}^{\top} = \begin{bmatrix} (P_{u}^{-1}\bar{u}_{0|k}^{*}-t_{u})^{\top}, \cdots, (P_{u}^{-1}\bar{u}_{N-1|k}^{*}-t_{u})^{\top} \end{bmatrix}^{\top}$} is a local (resp. global) minimizer of $\textbf{P-1}$. 
\end{lemma}
		
\begin{IEEEproof}
	With the input and output transformations in \eqref{Eq:uy_bar}, the cost term transformations in \eqref{Eq:paraAffine}, and the definitions of $J(\cdot, \cdot)$ and $\bar{J}(\cdot, \cdot)$, it can be shown that 
	\begin{equation} \label{Eq:J}
		J(y, u) = \bar{J}(\bar{y}, \bar{u}) + \varrho,
	\end{equation}
	where $\varrho = {\small\sum_{k=t}^{t+N-1}\left( L_{y}^{\top}\bar{Q}L_{y} + L_{u}^{\top}\bar{R}L_{u} \right) \in \mathbb{R}}$ is a constant. We now use proof by contradiction. % to facilitate the following proof. 
	Assume that $\left( \bar{y}^{*}, \bar{u}^{*} \right)$ is the global minimizer of optimization problem \eqref{Eq:AdaptedDeePCAffine}, and $\left(y^{*}, u^{*}\right)$ is the corresponding sequence of $\left(\bar{y}^{*}, \bar{u}^{*}\right)$ under the inverse affine maps $\left( P_{y}, L_{y} \right)$, $\left( P_{u}, L_{u} \right)$, i.e., 
	\[
	\begin{aligned}
		y^{*}& = \begin{bmatrix}
			y^{*}(t) 
			\\
			\vdots
			\\
			y^{*}(t+N-1)
		\end{bmatrix} = \begin{bmatrix}
			P_{y}^{-1}(y^{*}(t) - L_{y})
			\\
			\vdots
			\\
			P_{y}^{-1}(y^{*}(t+N-1) - L_{y})
		\end{bmatrix},
		\\
		u^{*}& = \begin{bmatrix}
			u^{*}(t)
			\\
			\vdots
			\\
			u^{*}(t+N-1)
		\end{bmatrix} = \begin{bmatrix}
			P_{u}^{-1}(u^{*}(t) - L_{u})
			\\
			\vdots
			\\
			P_{u}^{-1}(u^{*}(t+N-1) - L_{u})
		\end{bmatrix}.
	\end{aligned}
	\] 
	As $\left( \bar{y}^{*}, \bar{u}^{*} \right)$ is a trajectory of system \eqref{Eq:DT_TrafficModelAffine} and satisfies the constraints in \eqref{Eq:AdaptedDeePCAffine}, it can be confirmed that $\left(y^{*}, u^{*}\right)$ is a trajectory of system \eqref{Eq:DT_TrafficModel} and satisfies the constraints in \eqref{Eq:AdaptedDeePC}. We also assume that $\left(y^{*}, u^{*}\right)$ is not the global minimizer of problem \eqref{Eq:AdaptedDeePC}, and thus there exists an optimal sequence $\left(y^{**}, u^{**}\right)$ (other than $\left(y^{*}, u^{*}\right)$) such that
	\begin{equation} \label{Eq:J_ineq}
		J(y^{**}, u^{**}) < J(y^{*}, u^{*}).
	\end{equation}
	Let $\left(\bar{y}^{**}, \bar{u}^{**}\right)$ be the corresponding trajectory of $\left(y^{**}, u^{**}\right)$ under the forward affine maps $\left( P_{y}, L_{y} \right)$, $\left( P_{u}, L_{u} \right)$. According to \eqref{Eq:J}, \eqref{Eq:J_ineq} can be rewritten as 
	\begin{equation}
		\bar{J}(\bar{y}^{**}, \bar{u}^{**}) + \varrho < \bar{J}(\bar{y}^{*}, \bar{u}^{*}) + \varrho,
	\end{equation}
	which contradicts the assumption that $\left( \bar{y}^{*}, \bar{u}^{*} \right)$ is the global minimizer of optimization problem \eqref{Eq:AdaptedDeePCAffine}. 
 
 Therefore, if $\left( \bar{y}^{*}, \bar{u}^{*} \right)$ is the global minimizer of problem \eqref{Eq:AdaptedDeePCAffine}, then its inverse affine transformation, i.e., $\left(y^{*}, u^{*}\right)$, should be solution to the original problem \eqref{Eq:AdaptedDeePC}, indicating that these two problems are equivalent.
\end{IEEEproof}
		
\subsection{Privacy Preservation}\label{subsec_privacy}
		
We next discuss the privacy notion used in this paper. As mentioned in Section \ref{subsec_attackModel}, the attacker aims to infer the system state $x_{i}(k)$ and control input $u_{i}(k)$ of CAVs. Under the privacy-preserving architecture discussed above, the attacker will have access to $\bar{x}_{i}(k)$ and $\bar{u}_{i}(k)$ at each time step $k$, and we need to show that for any $\kappa\in \mathbb{Z}_{+}$, $x_{i, [1,\kappa]}$ and $u_{i, [1, \kappa]}$ cannot be identified from $\bar{x}_{i, [1, \kappa]}$ and $\bar{u}_{i, [1,\kappa]}$. According to \eqref{Eq:Affine}, we use  
\[
\left( x_{i, [1,\kappa]}, u_{i, [1, \kappa]} \right) \!\!\xRightarrow{\left( P_{x,i}, l_{x,i}, P_{u,i}, l_{u,i} \right)}\!\! \left( \bar{x}_{i, [1,\kappa]}, \bar{u}_{i, [1, \kappa]} \right), \; \forall i \!\in \! \Omega_{C}
\] 
to denote that $\left( \bar{x}_{i, [1,\kappa]}, \bar{u}_{i, [1, \kappa]} \right)$ is the transformed trajectory of $\left( x_{i, [1,\kappa]}, u_{i, [1, \kappa]} \right)$ under the affine maps $\left( P_{x,i}, l_{x,i} \right)$ and $\left( P_{u,i}, l_{u,i} \right)$. 

For any feasible state sequence $\bar{x}_{i, [1,\kappa]}$ and input sequence $\bar{u}_{i, [1,\kappa]}$ received by the attacker, the set $\Delta(\bar{x}_{i, [1,\kappa]}, \bar{u}_{i,[1,\kappa]})$ is defined as
%\begin{equation}
\begin{align*}
	&\Delta(\bar{x}_{i, [1,\kappa]}, \bar{u}_{i, [1,\kappa]}) = \lbrace x_{i, [1,\kappa]}, u_{i, [1,\kappa]}: \exists \left( P_{x,i}, l_{x,i}, P_{u,i}, l_{u,i} \right) %\nonumber
	\\
	&\text{s.t.} \; \left( x_{i, [1,\kappa]}, u_{i, [1, \kappa]} \right) \!\!\xRightarrow{\left( P_{x,i}, l_{x,i}, P_{u,i}, l_{u,i} \right)}\!\! \left( \bar{x}_{i, [1,\kappa]}, \bar{u}_{i, [1, \kappa]} \right) \rbrace.
\end{align*}
%\end{equation}
Essentially, the set $\Delta(\bar{x}_{i, [1,\kappa]}, \bar{u}_{i, [1,\kappa]})$ includes all possible values of $\left( x_{i, [1,\kappa]}, u_{i, [1,\kappa]} \right)$ that can be transformed into $\left( \bar{x}_{i, [1,\kappa]}, \bar{u}_{i, [1,\kappa]} \right)$ with corresponding affine maps $\left( P_{x,i}, l_{x,i}, P_{u,i}, l_{u,i} \right)$. 
		
\begin{definition} [$\infty$-Diversity] \label{def_privacy}
	The privacy of the actual system state $x_{i, [1,\kappa]}$ and input $u_{i, [1,\kappa]}$ of CAVs is preserved if the cardinality of the set $\Delta(\bar{x}_{i, [1,\kappa]}, \bar{u}_{i, [1,\kappa]})$ is infinite for any feasible state sequence $\bar{x}_{i, [1,\kappa]}$ and input sequence $\bar{u}_{i, [1,\kappa]}$.  
\end{definition}
		
The $\infty$-Diversity privacy definition requires that there are infinite sets of $\left( x_{i,[1,\kappa]}, u_{i,[1,\kappa]} \right)$ and $\left( P_{x,i}, l_{x,i}, P_{u,i}, l_{u,i} \right)$ that can generate the same $\left( \bar{x}_{i, [1,\kappa]}, \bar{u}_{i, [1,\kappa]} \right)$ received by the attacker. As a result, it is impossible for the attacker to use $\left( \bar{x}_{i, [1,\kappa]}, \bar{u}_{i, [1,\kappa]} \right)$ to infer the actual system state and input information.
		
\begin{remark}
	Definition \ref{def_privacy} of $\infty$-Diversity can be viewed as an extension to the $l$-diversity \cite{Machanavajjhala2007} that has been widely adopted in formal privacy analysis on attribute privacy of tabular. Essentially, $l$-diversity requires that there are at least $l$ different possible values for the privacy-sensitive data attributes, and a greater $l$ indicates greater indistinguishability.  \hfill $\square$   
\end{remark}
		
We next show that the proposed affine transformation mechanism can protect the privacy of CAVs based on Definition \ref{def_privacy}.
\begin{theorem} \label{theorem_privacy}
	Under the affine masking mechanism described in Section~\ref{subsec_affine}, the system state and control input of CAVs are $\infty$-Diversity, that is, the attacker cannot infer the actual system state $x_{i}(k)$ and input $u_{i}(k)$, $\forall i \in \Omega_{C}$.
\end{theorem}
		
\begin{IEEEproof}
	Based on Definition~\ref{def_privacy}, we prove Theorem~\ref{theorem_privacy} by showing that under the affine masking scheme, the cardinality of the set $\Delta(\bar{x}_{i, [1,\kappa]}, \bar{u}_{i, [1,\kappa]})$ is infinite. Specifically, given the sequence $\left( \bar{x}_{i, [1,\kappa]}, \bar{u}_{i, [1,\kappa]} \right)$ accessible to the attacker, for an arbitrary affine map $\left( P_{x,i}', l_{x,i}', P_{u,i}', l_{u,i}' \right)$ such that $ P_{x,i}'$ and $P_{u,i}'$ are invertible, a sequence $\left( x'_{i, [1,\kappa]}, u'_{i, [1,\kappa]} \right)$ can be uniquely determined based on $\left( \bar{x}_{i, [1,\kappa]}, \bar{u}_{i, [1,\kappa]} \right)$ by using $\left( P_{x,i}', l_{x,i}', P_{u,i}', l_{u,i}' \right)$ as an inverse mapping. As there exists infinitely many such affine maps $\left( P_{x,i}', l_{x,i}', P_{u,i}', l_{u,i}' \right)$, there exist infinitely many $\left( x'_{i,[1,\kappa]}, u'_{i,[1,\kappa]} \right)$ such that via proper affine transformations, the attacker will receive the same accessed information: $\left( \bar{x}_{i,[1,\kappa]}, \bar{u}_{i,[1,\kappa]} \right)$, which indicates that the cardinality of the set $\Delta(\bar{x}_{i, [1,\kappa]}, \bar{u}_{i, [1,\kappa]})$ is infinite.
\end{IEEEproof}
		
\begin{remark}
	Different from the conventional encryption-based techniques \cite{Darup2017CSL,Alexandru2018CDC}, the proposed affine masking-based privacy-preserved scheme does not involve complicated encryption and decryption procedure, and thus is light-weight in computation and can be easily implemented in a mixed traffic system. Furthermore, the affine masking mechanism can provide strong privacy protection such that the eavesdropper cannot even approximately estimate the interested information $x_{i}(k)$ and $u_{i}(k)$ via the exchanged information between the vehicle system and the central unit. \hfill $\square$
\end{remark}
		
\section{Numerical Experiments} \label{sec_perfEva}
\begin{figure}[t]
	\centering
	\includegraphics[scale=0.46]{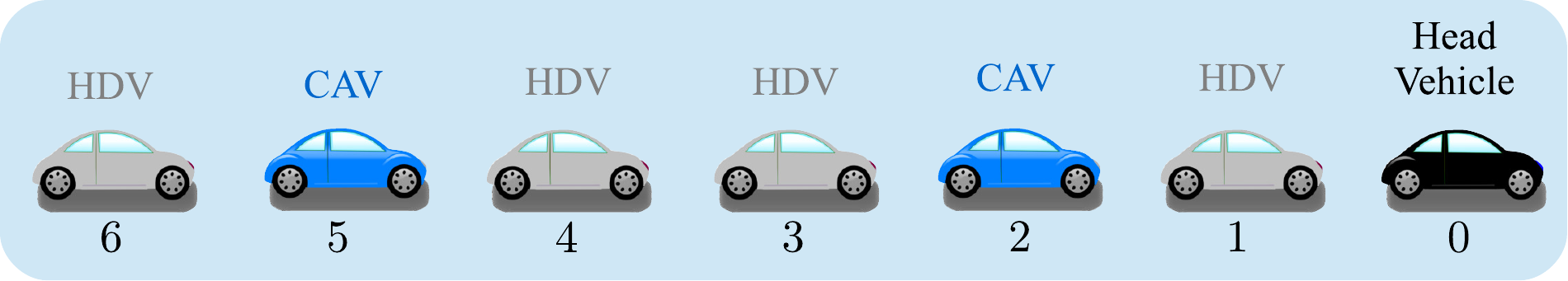}
	\caption{Mixed traffic system for simulation.}
	\label{Fig:VehicleSystem}
\end{figure}

In this section, we perform numerical simulations to validate the efficacy of the proposed privacy-preserving \method{DeeP-LCC}. Our implementation is adapted from the open-source code at \url{https://github.com/soc-ucsd/DeeP-LCC}. As shown in Fig.~\ref{Fig:VehicleSystem}, we consider a mixed traffic system consisting of two CAVs, four HDVs, and one head vehicle. The CAV and HDV indices are $\Omega_{C}=\left\{ 2, 5 \right\}$ and $\Omega_{H}=\left\{1, 3, 4, 6 \right\}$, respectively. As discussed in Section~\ref{sec_privacyPreserved}, there exists a communication network between the central unit and vehicles. The central unit collects the vehicle data and then computes the control actions for the CAVs.  

\begin{table}[t]
\begin{center}
\caption{Parameter Setup for \method{DeeP-LCC} \eqref{Eq:AdaptedDeePC}}\label{Tb:ParameterSetup}
\begin{threeparttable}
	\setlength{\tabcolsep}{2mm}{
		\begin{tabular}{c c c}
		\hline
		\hline
		Notation & Meaning & Value \\					
		\hline
		$N$ & Length for future data horizon & 30 \\
		$T_{\mathrm{ini}}$ & Length for past data horizon & 15 \\
		$w_{s}$ & Weight coefficient for spacing error & 0.5 \\
	    $w_{v}$ & Weight coefficient for velocity error & 1 \\
		$w_{u}$ & Weight coefficient for control input & 0.1 \\
		$\tilde{s}_{\mathrm{min}}$ [$\mathrm{m}$] & Lower bound of spacing error & -15 \\
		$\tilde{s}_{\mathrm{max}}$ [$\mathrm{m}$] & Upper bound of spacing error & 20 \\
		$\tilde{v}_{\mathrm{min}}$ [$\mathrm{m/s}$] & Lower bound of velocity error & -30 \\
		$\tilde{v}_{\mathrm{max}}$ [$\mathrm{m/s}$] & Upper bound of velocity error & 30 \\
		$a_{\mathrm{min}}$ [$\mathrm{m/s^2}$] & Lower bound of acceleration & -5 \\
		$a_{\mathrm{max}}$ [$\mathrm{m/s^2}$] & Upper bound of acceleration & 2 \\
		\hline
		\hline
	\end{tabular}}
\end{threeparttable}
\end{center}
\vspace{-4mm}
\end{table}

The OVM model~\cite{Bando1995PRE} is used to describe the behavior of HDVs, and noises with uniform distribution of $\mathbb{U}\left[-0.3, 0.3\right]$ are added to the HDVs' model. Note that the OVM model is only used to update the state of HDVs and is not utilized by the developed control scheme. Both the original \method{DeeP-LCC} and the proposed privacy-preserving \method{DeeP-LCC} are implemented in the numerical simulations. We follow a similar procedure introduced in \cite{Wang2022arXiv} to collect the offline data $\left(u^{\mathrm{d}}, \epsilon^{\mathrm{d}}, y^{\mathrm{d}}\right)$ with the sampling interval chosen as $\Delta t=0.05$ s. For the \method{DeeP-LCC} \eqref{Eq:AdaptedDeePC}, the parameter setup is listed in Table~\ref{Tb:ParameterSetup}.
%the weighting factors are selected as $w_{s} = 0.5$, $w_{v} = 1$, and $w_{u} = 0.1$. 
%The lower and upper bounds of output signal of the CAVs are chose as $y_{\mathrm{min}} = \begin{bmatrix} -15, -30 \end{bmatrix}^{\top}$, $y_{\mathrm{max}} = \begin{bmatrix} 20, 30 \end{bmatrix}^{\top}$, while the acceleration limits of the CAVs are set as $a_{\mathrm{min}} = -5$ ${\color{blue}\mathrm{m/s^2}}$, $a_{\mathrm{max}} = 2$ ${\color{blue}\mathrm{m/s^2}}$. 
%The lower and upper bounds for the spacing of the CAVs are set to $\tilde{s}_{\mathrm{min}} = -15$ ${\color{blue}\mathrm{m}}$, $\tilde{s}_{\mathrm{max}} = 20$ ${\color{blue}\mathrm{m}}$, the boundaries for the velocity error of each vehicle are set to $\tilde{v}_{\mathrm{min}} = -15$ ${\color{blue}\mathrm{m/s}}$, $\tilde{v}_{\mathrm{max}} = 20$ ${\color{blue}\mathrm{m/s}}$, and the acceleration limits of the CAVs are set as $a_{\mathrm{min}} = -5$ ${\color{blue}\mathrm{m/s^2}}$, $a_{\mathrm{max}} = 2$ ${\color{blue}\mathrm{m/s^2}}$. 
For the proposed privacy-preserving \method{DeeP-LCC}, CAVs exploit affine transformation maps \eqref{Eq:Affine} to mask their actual state and input. The affine transformation maps for CAVs (recall that $\Omega_{C}=\left\{ 2, 5 \right\}$) are chosen as
\[
\begin{aligned}
	P_{x,2} &= \begin{bmatrix}
	\cos(\frac{\pi}{4}) & -\sin(\frac{\pi}{4}) \\			
	\sin(\frac{\pi}{4}) & \cos(\frac{\pi}{4})
	\end{bmatrix}, l_{x,2}=\begin{bmatrix}
	5 \\ 3
	\end{bmatrix}, \\
	P_{u,2} &= -1.5, l_{u,2} = 1,
\end{aligned}
\]
and
\[
\begin{aligned}
	P_{x,5} &= \begin{bmatrix}
		\cos(\frac{8\pi}{9}) & -\sin(\frac{8\pi}{9}) \\
		\sin(\frac{8\pi}{9}) & \cos(\frac{8\pi}{9})
	\end{bmatrix}, l_{x,5}=\begin{bmatrix}
		5 \\ 3
	\end{bmatrix}, \\
	P_{u,5} &= 1.5, l_{u,5} = -1.
\end{aligned}
\] 
Furthermore, based on the aforementioned affine maps and the parameter setup for \method{DeeP-LCC}, the parameters used to formulate the privacy-preserving \method{DeeP-LCC} problem \ref{Eq:AdaptedDeePCAffine} can be obtained (see Section~\ref{subsec_affine} and \ref{subsec_reformulation}). 
		
\subsection{Scenario A: Comprehensive Acceleration and Deceleration}
Motivated by the New European Driving Cycle (NEDC)~\cite{DieselNet} and the experiments in \cite{Wang2022arXiv}, a comprehensive acceleration and deceleration scenario is designed to validate the capability of the proposed privacy-preserving \method{DeeP-LCC} in improving traffic performance. 
The NEDC combines ECE 15 with Extra Urban Driving Cycle (EUDC) to assess the fuel economy of vehicles. Similar to ECE 15 and EUDC, 
we design a driving trajectory for the head vehicle that incorporates acceleration and deceleration at different time periods. Both fuel consumption and velocity errors for the vehicles are considered to quantify traffic performance. 

More specifically, for the $i$-th vehicle, the fuel consumption rate $f_{i}$ (mL/s) is calculated based on an instantaneous model in~\cite{Bowyer1985}, which is given by
\[
f_{i} = 
\begin{cases}
0.444 + 0.090 R_{i}v_{i} + \left[ 0.054a_{i}^{2}v_{i} \right]_{a_{i}>0}, \quad \text{if } R_{i}>0,
\\
0.444, \qquad \qquad \qquad \qquad \qquad \qquad \qquad \;\; \text{if } R_{i}\le 0,
\end{cases}
\] 
with $R_{i}=0.333+0.00108v_{i}^{2}+1.200a_{i}$. Since the first HDV is not influenced by the CAVs, the total fuel consumption rate $f$ of the mixed traffic system is calculated by summing $f_{i}$ indexed from 2 to 6, i.e., $f = \sum_{i=2}^{6} f_{i}$. The average absolute velocity error (AAVE) is utilized to quantify velocity errors and is obtained by computing the average of $\frac{\left| v_{i}(t)-v_{0}(t) \right|}{v_{0}(t)}$ with respect to the simulation time and the vehicle number. Moreover, the original \method{DeeP-LCC} scheme and the proposed privacy-preserving \method{DeeP-LCC} under the Hankel and Page matrix structures are tested in this scenario. The number of columns in Hankel and Page matrices are selected as 900, which is sufficient for the non-parametric representation of the mixed traffic system with 2 CAVs and 4 HDVs. A standard output-feedback MPC is also tested in this scenario to facilitate a complete comparison. The MPC is designed based on the linearized mixed traffic system model~\eqref{Eq:DT_TrafficModel}, and its future time horizon, cost function, and constraints are the same as those of \method{DeeP-LCC}. Each method is carried out one time in this scenario, and the fuel consumption and AAVE indices are computed to evaluate the performance.
\begin{figure}[!t]
	\setlength{\abovecaptionskip}{0pt}
	\centering
	\subfigure[] {\label{fig_vel_HDV_nedc}
	\includegraphics[width=0.5\textwidth]{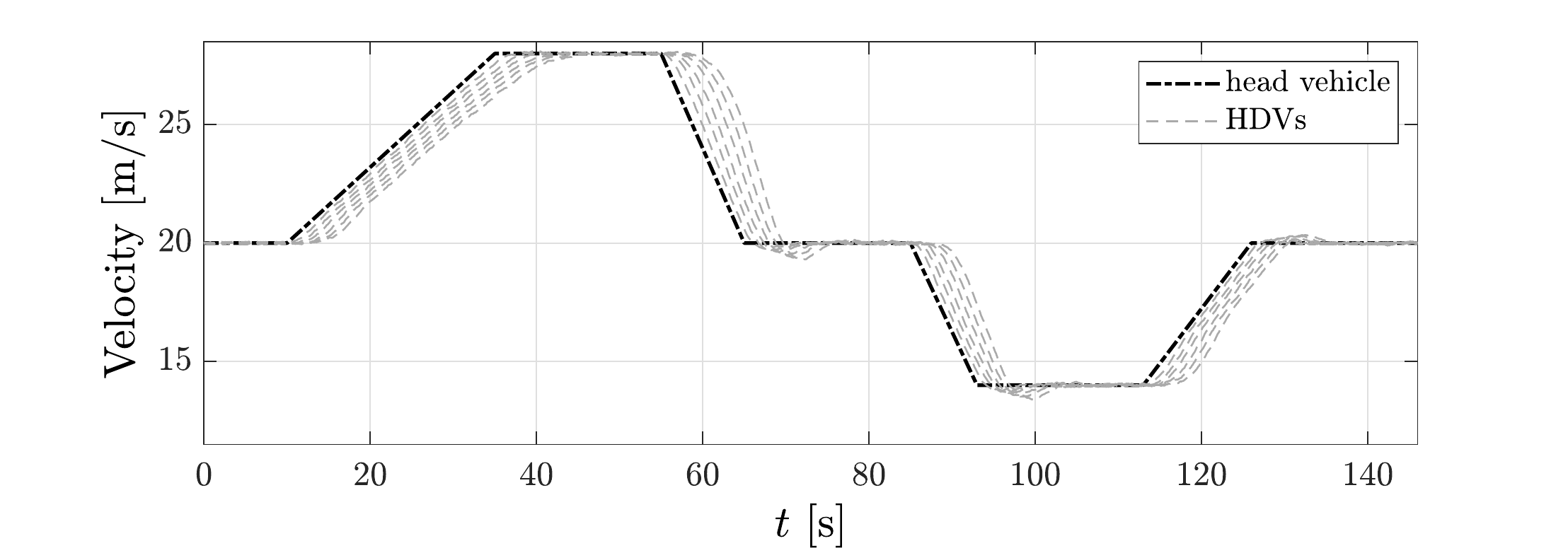}
	}
	\subfigure[] {\label{fig_vel_DeeP-LCC_nedc}
	\includegraphics[width=0.5\textwidth]{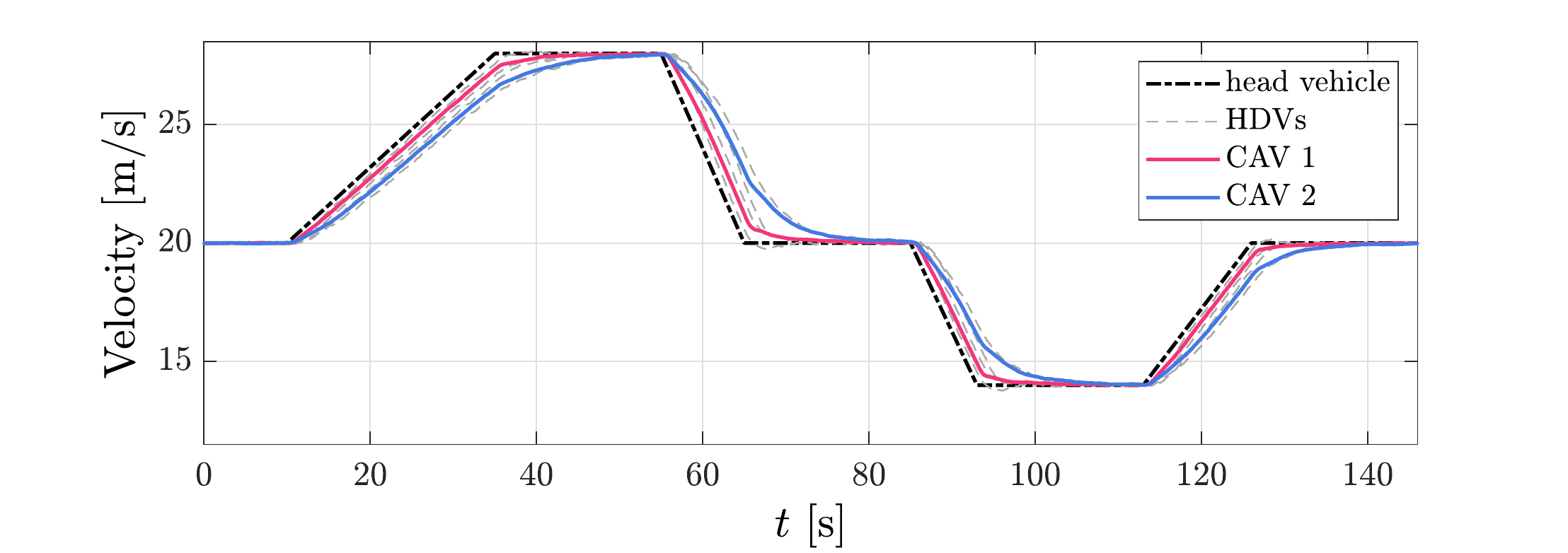}
	}
	\subfigure[]{\label{fig_vel_PrivacyDeeP-LCC_nedc}
	\includegraphics[width=0.5\textwidth]{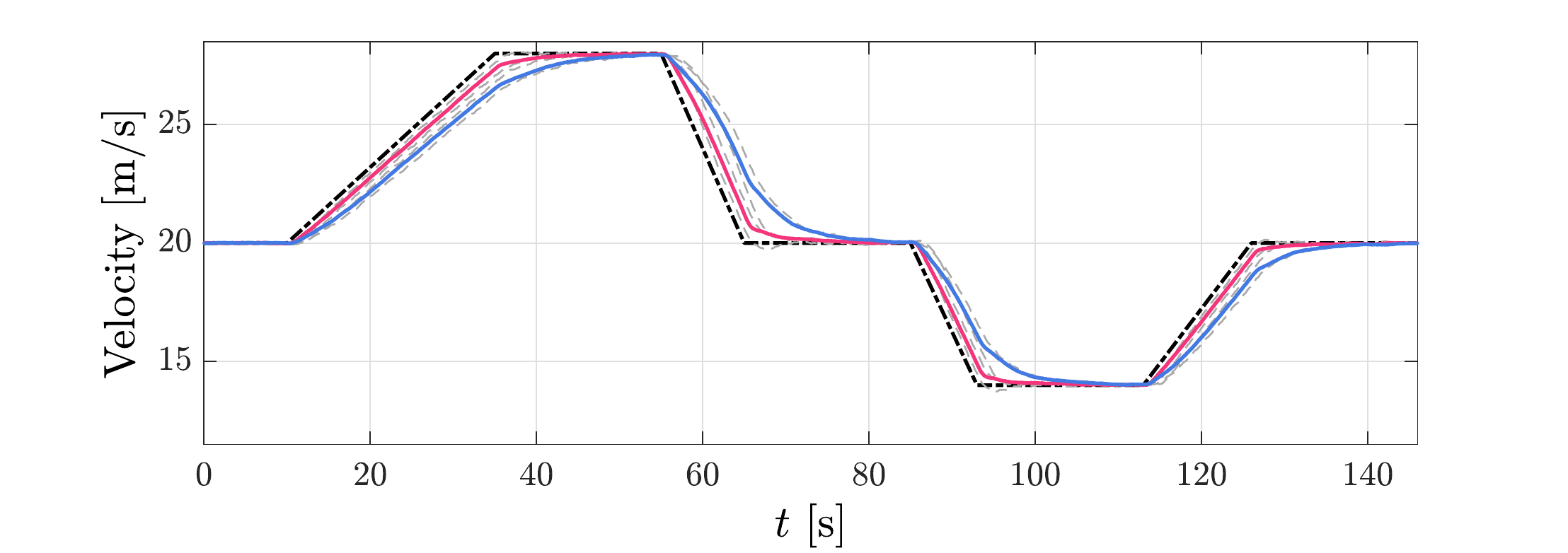}
	}
	\caption{Velocity profiles of the mixed traffic system under Scenario A. (a) All the vehicles are HDVs. (b) \method{DeeP-LCC} under the Hankel matrix structure is utilized to generate the control for CAVs. (c) Privacy-preserving \method{DeeP-LCC} under the Hankel matrix structure is utilized to generate the control for CAVs.}
	\label{fig_velocity_nedc}
\end{figure}

The simulation results are shown in Fig.~\ref{fig_velocity_nedc}. Note that the velocity response profiles of MPC and Page matrix based \method{DeeP-LCC} schemes are quite similar to the ones of Hankel matrix based \method{DeeP-LCC} schemes, and hence they are omitted. As shown in Fig.~\ref{fig_velocity_nedc}, compared to the case with all HDVs, both the original \method{DeeP-LCC} and the proposed privacy-preserving \method{DeeP-LCC} can effectively mitigate velocity fluctuations, leading to a smoother mixed traffic flow. Table~\ref{Tb:NEDC} presents the fuel consumption and AAVE of different schemes. All \method{DeeP-LCC} approaches achieve comparable performance with the ideal MPC setting. Note that MPC relies on the nominal model (6) to facilitate the controller design (that is generally not available), while \method{DeeP-LCC} approaches directly utilize pre-collected trajectory data to generate control inputs. The explicit model for each HDV is generally unknown and hard to identify due to stochastic and uncertain human driving behavior, and thus MPC is not easily applicable to real-world deployment. In contrast, \method{DeeP-LCC} approaches can bypass system identification and accomplish similar performance with MPC based on pre-collected data, which might be easier to deploy for mixed traffic with enhanced communication technologies. 

In addition, it can be found from Table~\ref{Tb:NEDC} that the proposed privacy-preserving method still retains the advantages of the original \method{DeeP-LCC} in improving fuel economy and traffic smoothness. As revealed in Lemma 1, although the affine transformation mechanism is introduced to mask the actual system state and input signals, the resulting privacy-preserving \method{DeeP-LCC}~\eqref{Eq:AdaptedDeePCAffine} is equivalent to the original \method{DeeP-LCC}~\eqref{Eq:AdaptedDeePC}. Therefore, the proposed privacy protection scheme is shown not to induce degradation on control performance.
\begin{table}[t]
\begin{center}
	\caption{Fuel Consumption and Average Absolute Velocity Error (AAVE) in Scenario A}\label{Tb:NEDC}
	\begin{threeparttable}
		\setlength{\tabcolsep}{2mm}{
			\begin{tabular}{c c c}
				\hline
				\hline
				& Fuel Consumption [mL] & AAVE [$10^{-3}$] \\
				\hline
				All HDVs & 1569.62 & 27.34 \\
				MPC & 1536.98($\downarrow$2.08\%) & 24.50($\downarrow$10.38\%) \\
				DeeP-LCC (Hankel) & 1537.98($\downarrow$2.02\%) & 24.52($\downarrow$10.29\%) \\
				PP-DeeP-LCC (Hankel) & 1538.71($\downarrow$1.97\%) & 24.48($\downarrow$10.47\%) \\
				DeeP-LCC (Page) & 1537.66($\downarrow$2.07\%) & 24.48($\downarrow$10.45\%) \\
				PP-DeeP-LCC (Page) & 1538.73($\downarrow$1.97\%) & 24.50($\downarrow$10.39\%) \\
				\hline
				\hline
		\end{tabular}}
	\begin{tablenotes}
	\footnotesize
	\item[1] All the values have been rounded. PP-DeeP-LCC refers to privacy-preserving DeeP-LCC.
	\end{tablenotes}
	\end{threeparttable}
\end{center}
\vspace{-4mm}
\end{table}

\begin{table}[t]
\begin{center}
\caption{Variable Dimension and Computation Time in Scenario A}\label{Tb:NEDC_time}
\begin{threeparttable}
\setlength{\tabcolsep}{2mm}{
\begin{tabular}{c c c}
	\hline
	\hline
	& \begin{tabular}{@{}c@{}}Decision Variable \\ Dimension\end{tabular} & \begin{tabular}{@{}c@{}}Mean Computation \\ Time [ms]\end{tabular} \\
	\hline
	MPC & 60 & 3.58 \\
	DeeP-LCC (Hankel) & 900 & 26.31 \\
	PP-DeeP-LCC (Hankel) & 900 & 27.71 \\
	DeeP-LCC (Page) & 900 & 27.49 \\
	PP-DeeP-LCC (Page) & 900 & 27.91 \\
	\hline
	\hline
\end{tabular}}
\end{threeparttable}
\end{center}
\vspace{-4mm}
\end{table}

Both the original \method{DeeP-LCC}~\eqref{Eq:AdaptedDeePC} and the privacy-preserving \method{DeeP-LCC}~\eqref{Eq:AdaptedDeePCAffine} can be formulated into a standard quadratic program, and the dimension of their main decision variables $g$/$\bar{g}$ is identical to the column number of data matrix. In this simulation, both Hankel and Page matrices have the same column number of $900$, i.e., $g$, $\bar{g}\in \mathbb{R}^{900}$. For MPC, its main decision variable is the future control sequence, which has the size of $N\times m = 60$. Therefore, the online optimization complexity of \method{DeeP-LCC} approaches is higher than that of MPC. We run the MPC and \method{DeeP-LCC} approaches in a computer with Intel Core i7-12700K CPU and 16G RAM, and the computation time is concluded in Table~\ref{Tb:NEDC_time}, %. Due to the difference in optimization complexity, 
which confirms that the \method{DeeP-LCC} approaches require more computation time than the MPC. However, without further customization or optimization of the code, the mean computation time of \method{DeeP-LCC} approaches during this simulation is less than 30 ms, which is acceptable for %real-time 
deployment in the considered mixed traffic system. Improving the computational efficiency of \method{DeeP-LCC} for large-scale system is an interesting future direction. We refer the interested readers to~\cite{Zhang2022arXiv,Wang2022Distributed} for recent potential strategies based on dimension reduction and distributed optimization techniques.

\subsection{Scenario B: Highway Vehicle Trajectory}
\begin{figure}[t]
	%\vspace{1mm}
	\centering
	\includegraphics[width=0.3585\textwidth]{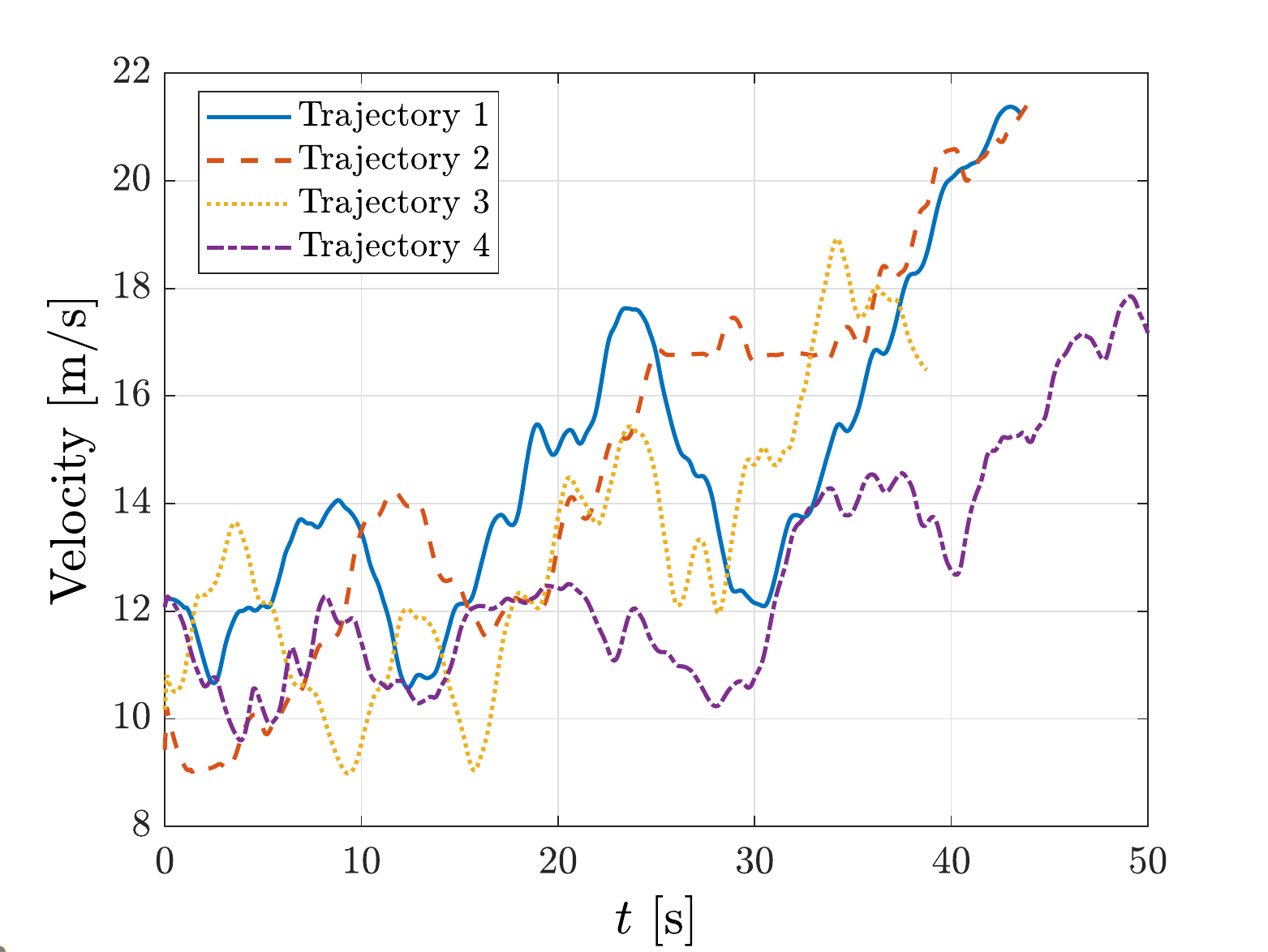}
	%\vspace{-3mm}
	\caption{Four vehicle trajectories from the NGSIM program. The trajectories are inputted as the head vehicle trajectory.}
	\label{fig_trajectory_NGSIM}
	%\vspace{-5mm}
\end{figure}

In this scenario, the real vehicle trajectory from the Next Generation SIMulation (NGSIM) program~\cite{NGSIM} is used to further validate the proposed privacy-preserving \method{DeeP-LCC} scheme. The NGSIM program provides comprehensive high-quality traffic data collected from two freeway segments and two arterial segments. We use the traffic data collected on a freeway segment of US-101 to facilitate the simulation. In particular, four vehicle trajectories (i.e., No. 2, 22, 42, and 48) between 8:05 a.m. to 8:20 a.m. in the US-101 dataset are extracted and inputted as the head vehicle trajectory of the mixed traffic system. Fig.~\ref{fig_trajectory_NGSIM} shows the velocity profile of these four trajectories. All trajectories include significant oscillating procedure and thus are selected to validate the effectiveness of the proposed scheme in terms of fuel economy and traffic smoothness.
	
\begin{figure}[!t]
	\setlength{\abovecaptionskip}{0pt}
	\centering
	\subfigure[] {\label{fig_fuelConsumption_ngsim}
		\includegraphics[width=0.5\textwidth]{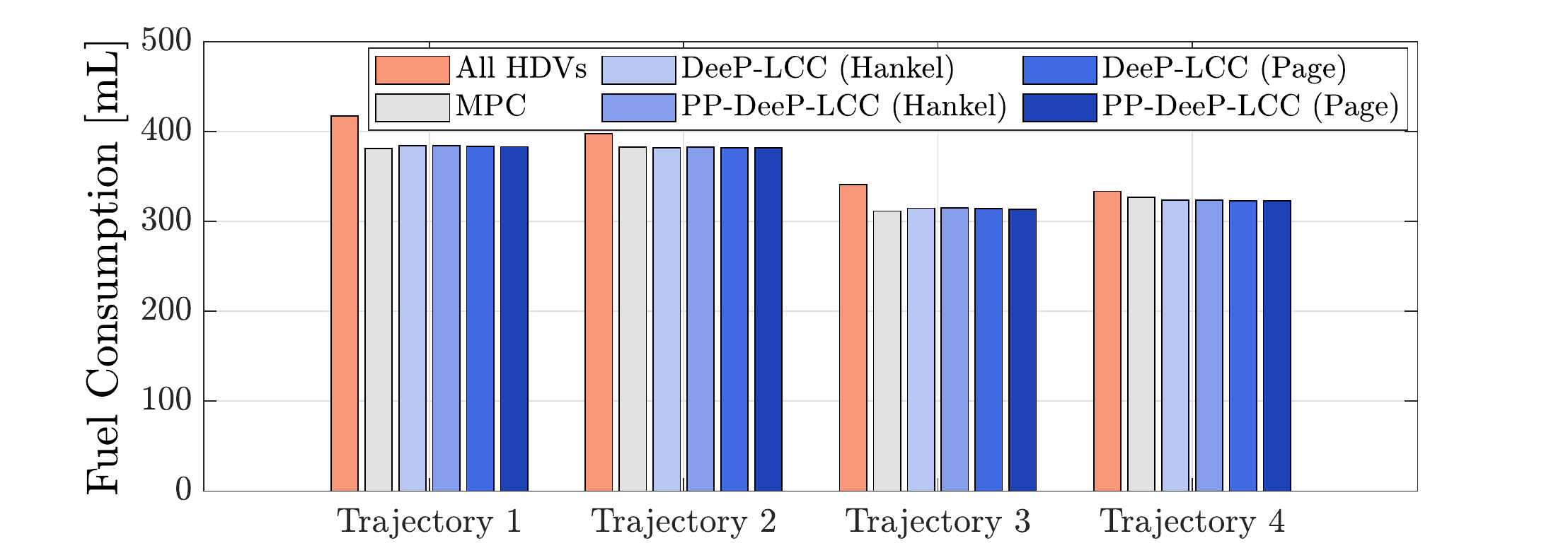}
	}
	%\vspace{-2mm}
	\subfigure[] {\label{fig_AAVE_ngsim}
		\includegraphics[width=0.5\textwidth]{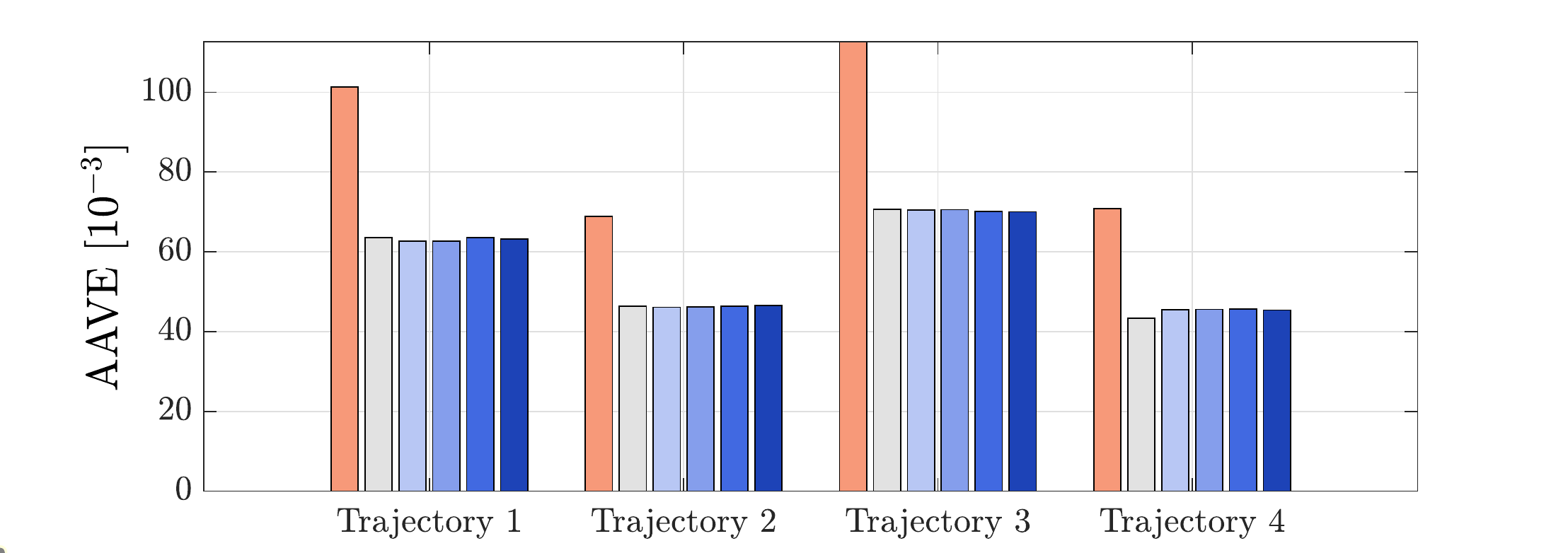}
	}
	\caption{Performance improvement of MPC and \method{DeeP-LCC} approaches compared with the case where all the vehicles are HDVs under Scenario B: (a) Fuel Consumption; and (b) AAVE.}
	\label{fig_comparison_ngsim}
\end{figure}

MPC and all \method{DeeP-LCC} related schemes are tested in this scenario, and the fuel consumption and AAVE indices are utilized to evaluate the performance. 
%The reduction rate of fuel consumption and AAVE by MPC and \method{DeeP-LCC} approaches with respect to the case where all the vehicles are HDVs is illustrated in Fig.~\ref{fig_comparison_ngsim}.
The performance improvement of MPC and \method{DeeP-LCC} approaches compared with the case where all the vehicles are HDVs is illustrated in Fig.~\ref{fig_comparison_ngsim}.
It is evident that MPC and all \method{DeeP-LCC} approaches substantially improve the fuel economy and traffic smoothness. Moreover, the improved traffic behavior under the proposed privacy-preserving \method{DeeP-LCC} is quite similar to that of the original \method{DeeP-LCC}, which indicates that the developed privacy protection method does not sacrifice traffic control efficacy for privacy.

\subsection{Scenario C: Emergency Braking}
We finally use a braking scenario to show that the proposed method can protect the privacy of CAVs against the attacker. In this scenario, the head vehicle takes a sudden emergency brake with maximum deceleration, then maintains the low velocity for a while, and finally accelerates to the original velocity. This is a typical emergency situation in real-world traffic, and the control actions of CAVs should provide strict safety guarantees to prevent rear-end collision.
\begin{figure}[!t]
	\setlength{\abovecaptionskip}{0pt}
	\centering
	\subfigure[] {\label{fig_vel_HDV}
		\includegraphics[width=0.239\textwidth]{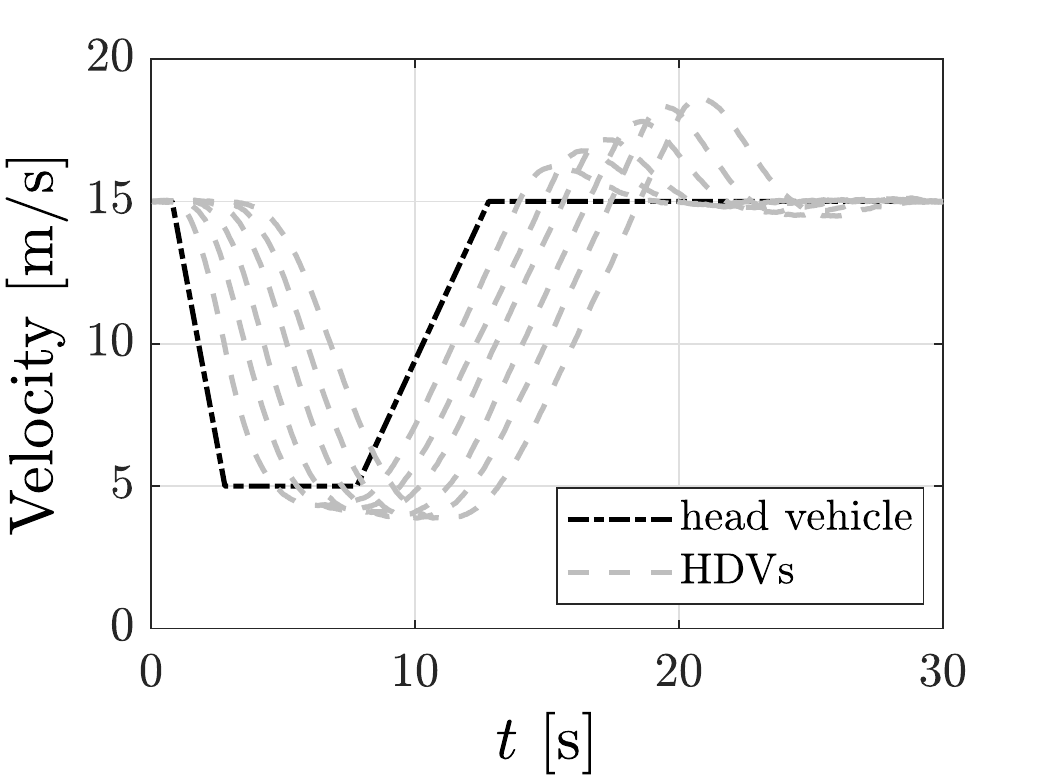}		
	}
	\hspace{0 in}
	\subfigure[] {\label{fig_vel_DeeP-LCC}
		\includegraphics[width=0.239\textwidth]{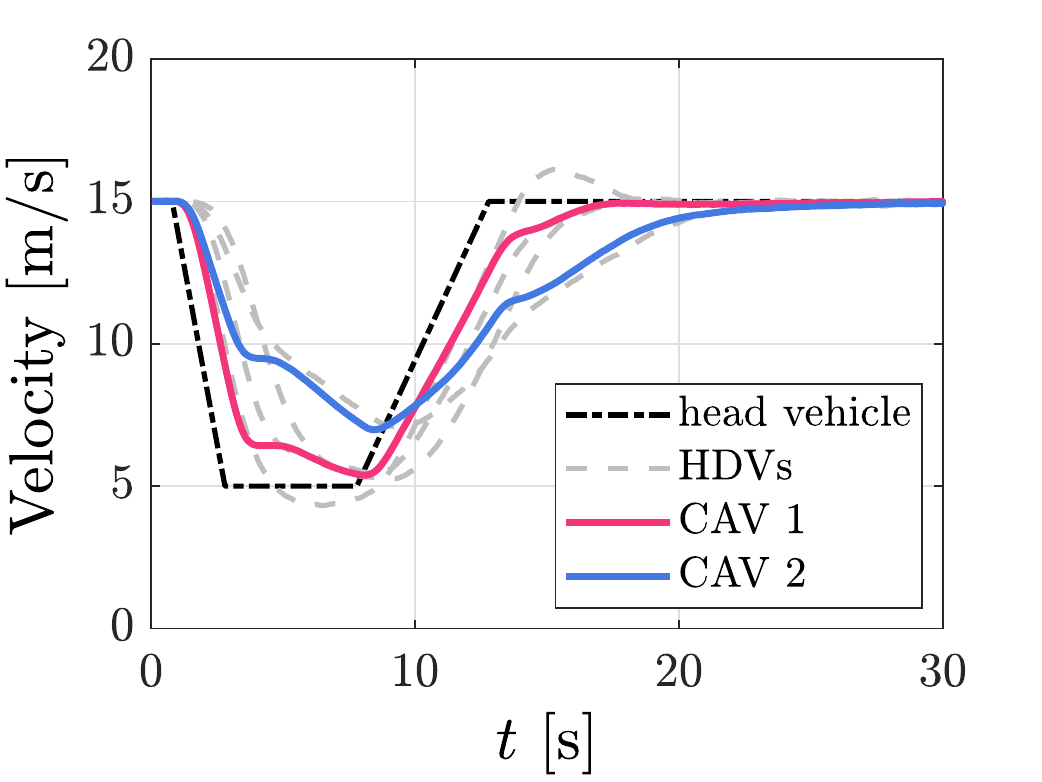}
	}
	\hspace{-0.22 in}
	\subfigure[] {\label{fig_vel_PrivacyDeeP-LCC}
		\includegraphics[width=0.239\textwidth]{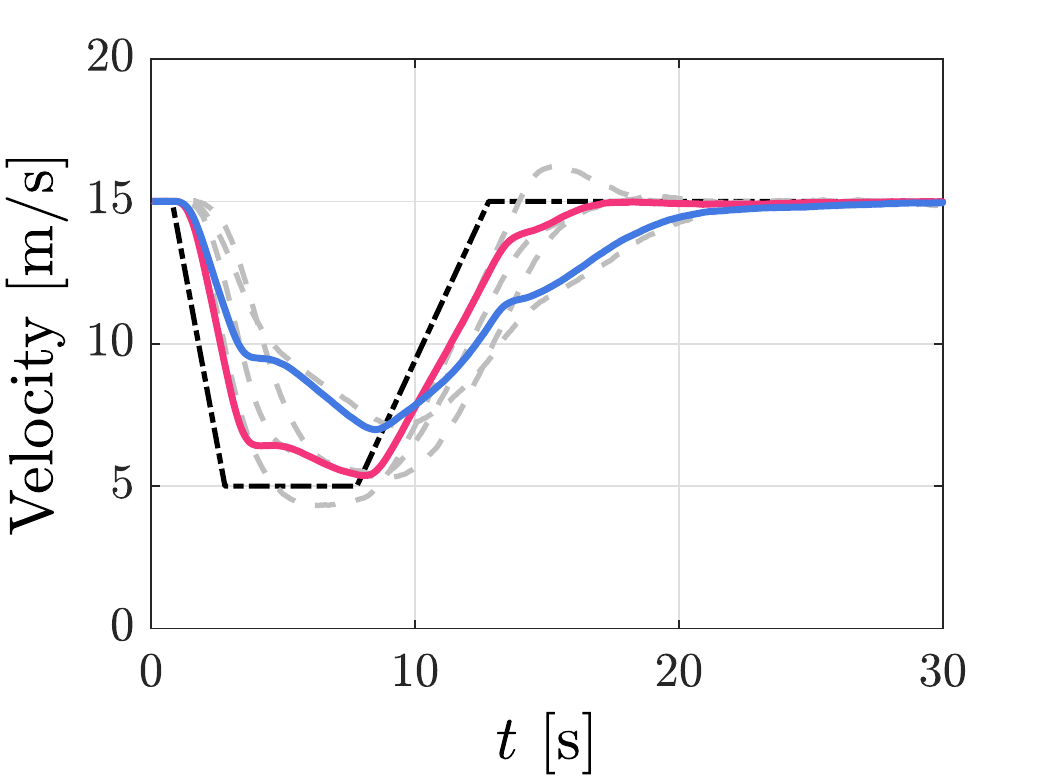}
	}
	\caption{Velocity profiles of the mixed traffic system under Scenario C. (a) All the vehicles are HDVs. (b) \method{DeeP-LCC} under the Page matrix structure is utilized to generate the control for CAVs. (c) Privacy-preserving \method{DeeP-LCC} under the Page matrix structure is utilized to generate the control for CAVs.}
	\label{fig_velocity}
\end{figure}	
\begin{figure}[!t]
	\setlength{\abovecaptionskip}{0pt}
	\centering
	\subfigure[] {\label{fig_stateInput_PrivacyDeeP-LCC}
		\includegraphics[width=0.239\textwidth]{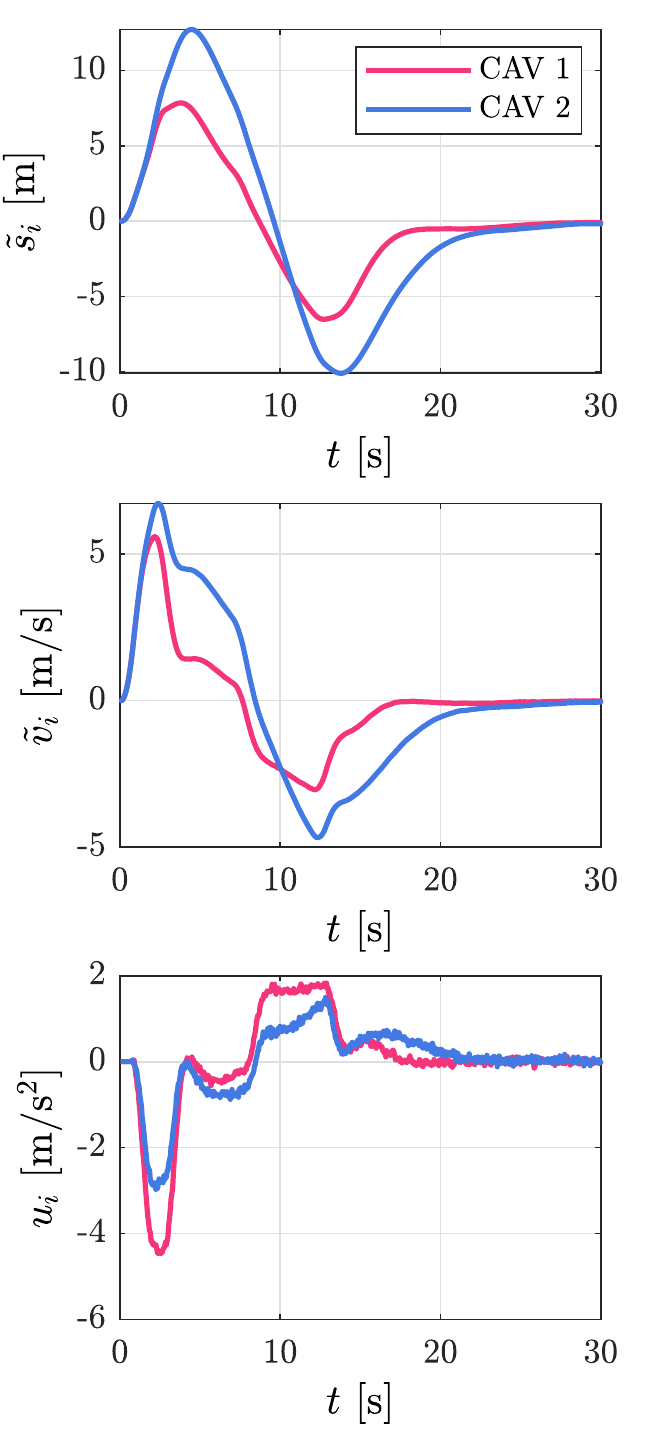}
	}
	\hspace{-0.22 in}
	\subfigure[] {\label{fig_stateInputMask_PrivacyDeeP-LCC}
		\includegraphics[width=0.239\textwidth]{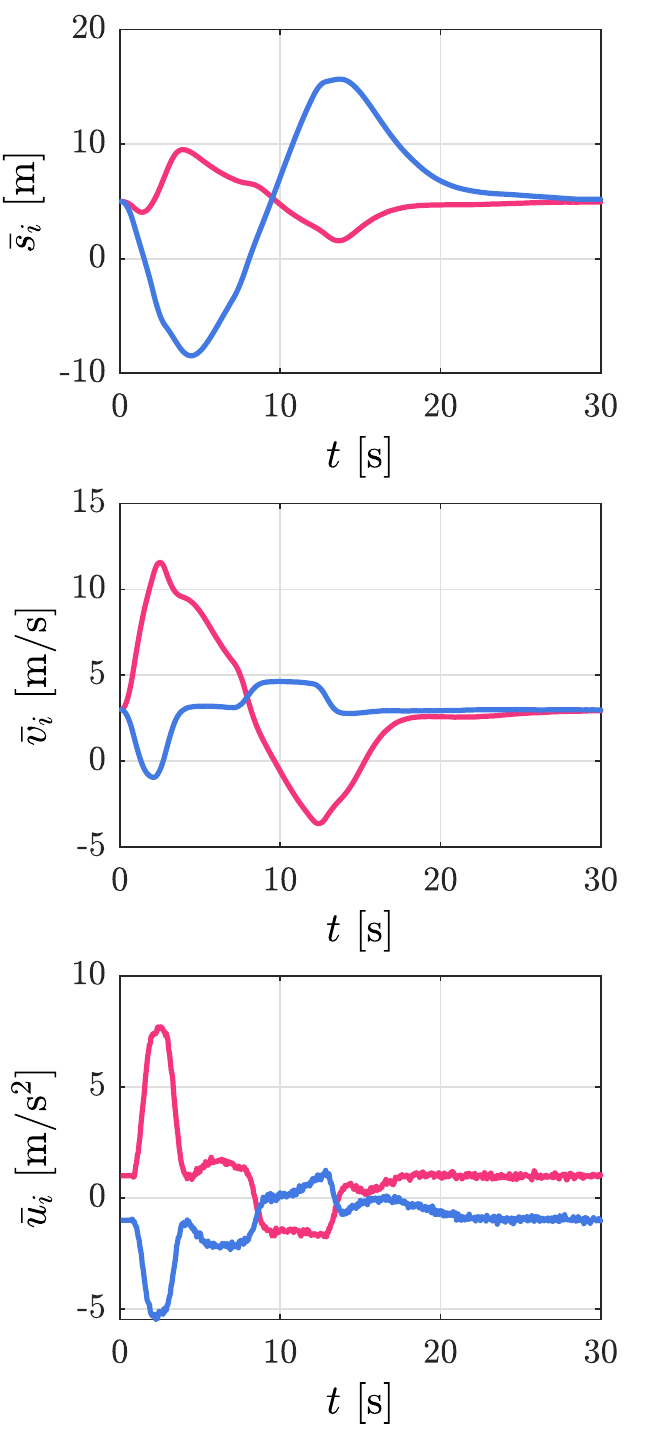}
	}
	\caption{Simulation results with privacy-preserving \method{DeeP-LCC} under Scenario C: (a) actual state and input information of CAVs, (b) state and input information exchanged between the CAVs and the central unit.}
	\label{fig_stateInput}
\end{figure}
		
The simulation results are presented in Figs.~\ref{fig_velocity} and \ref{fig_stateInput}. It can be found from Fig.~\ref{fig_velocity} that when all the vehicles are HDVs, their velocities have a large fluctuation in response to the head vehicle's braking perturbation. This velocity fluctuation will greatly increase fuel consumption and collision risk. By comparison, under the \method{DeeP-LCC} strategy and the proposed privacy-preserving \method{DeeP-LCC}, the CAVs show similar response patterns to mitigate velocity perturbations and smooth the mixed traffic flow. More precisely, when the head vehicle begins to brake (see the time period before 8 s), the CAVs decelerate immediately to adjust the relative distance from the preceding vehicle; when the head vehicle starts to return to the original velocity (see the time period after 8 s), the CAVs accelerate gradually. The similar motion pattern under these two methods implies that the affine transformation mechanism can retain the control performance as the original \method{DeeP-LCC}. 

Recall that for the original \method{DeeP-LCC} architecture introduced in Section~\ref{subsec_attackModel}, each CAV needs to exchange real-time information with the central unit, which will directly leak privacy-sensitive messages $x_{i}(k)$ and $u_{i}(k)$ to the external eavesdropper or the honest-but-curious central unit. To avoid privacy leakage, we design a masking mechanism to modify the exchanged messages between the CAVs and the central unit. Fig.~\ref{fig_stateInput} shows the actual state and input information of CAVs (i.e., $x_{i}(k)=\begin{bmatrix}
\tilde{s}_{i}(k), \tilde{v}_{i}(k)
\end{bmatrix}^{\top}, u_{i}(k), i\in \Omega_{C}$), and the corresponding modified information exchanged between the CAVs and the central unit (i.e., $\bar{x}_{i}(k)=\begin{bmatrix}
\bar{s}_{i}(k), \bar{v}_{i}(k)
\end{bmatrix}^{\top}, \bar{u}_{i}(k), i\in \Omega_{C}$) under the privacy-preserving \method{DeeP-LCC}. It can be observed that the information exchanged between the CAVs and the central unit is quite different from the actual one.
%Meanwhile as shown in Fig.~\ref{fig_stateInput}, under the privacy-preserving \method{DeeP-LCC}, the exchanged information between the CAVs and the central unit, i.e., $\bar{x}_{i}(k)=\begin{bmatrix} \bar{s}_{i}(k), \bar{v}_{i}(k) \end{bmatrix}^{\top}, \bar{u}_{i}(k), i\in \Omega_{C}$, is quite different from the actual one, i.e., $x_{i}(k)=\begin{bmatrix} \tilde{s}_{i}(k), \tilde{v}_{i}(k) \end{bmatrix}^{\top}, u_{i}(k), i\in \Omega_{C}$. 
This indicates that the proposed affine transformation mechanism can conceal the actual state and input of CAVs, which makes the external eavesdropper or the honest-but-curious central unit unable to infer $x_{i}(k)$ and $u_{i}(k)$.	
		
%\subsection{Influences of Hankel and Page matrix structures}
		
\begin{figure}[t]
	\centering
	\includegraphics[width=0.3585\textwidth]{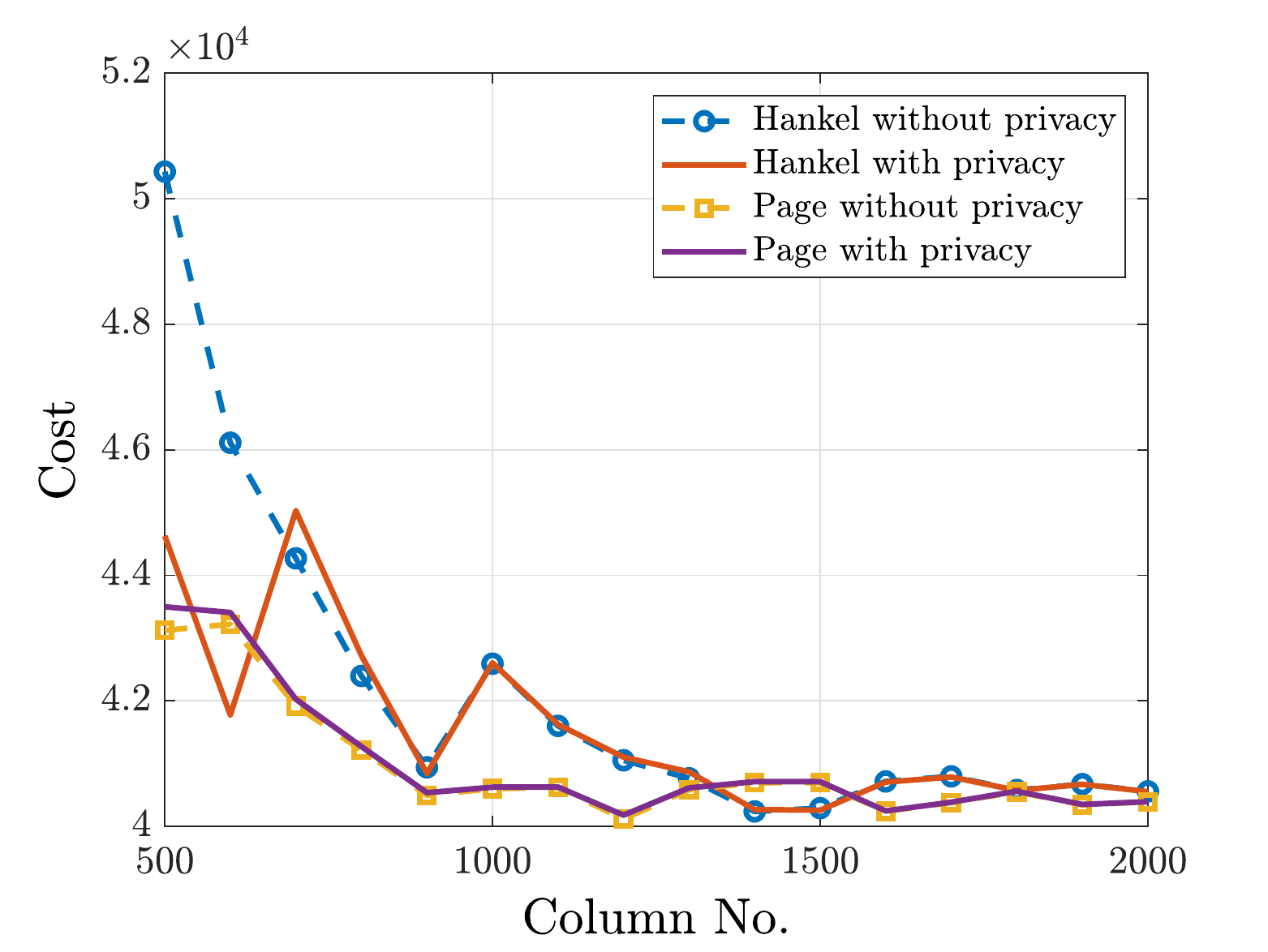}
	\caption{Cost evolution under different number of columns in data matrix.}
	\label{fig_comparison_cost}
\end{figure}

As discussed in Remark~\ref{remark_Page}, the key difference between the Hankel matrix and the Page matrix is that there are no repeated entries in the latter.
We conduct simulation to evaluate the performance of Hankel and Page matrix structures and analyze the effect of this difference. Specifically, with number of columns in data matrix ranging from 500 to 2000, the original \method{DeeP-LCC} without privacy protection and the proposed privacy-preserving \method{DeeP-LCC} are tested in Scenario B. The quadratic cost defined in \eqref{Eq:AdaptedDeePC} is calculated to evaluate the performance. Fig.~\ref{fig_comparison_cost} shows the cost of the mixed traffic system with different number of columns in Hankel and Page matrices. It can be seen that when the number of columns is less than 1300, compared to the Hankel matrix based \method{DeeP-LCC}, the Page matrix based ones show less cost volatility. Since none of the entries in the Page matrix are repeated, the Page matrix can store more data than the Hankel matrix with the same number of columns. More data samples generally contain more information about the system, hence the Page matrix based \method{DeeP-LCC} could be more efficient in representing the system via the non-parametric manner, resulting in lower cost fluctuations. When the number of columns is larger than 1300, both these two matrix structures store adequate data, and thus the cost of Hankel matrix based \method{DeeP-LCC} and Page matrix based \method{DeeP-LCC} is stable around a constant.
%It can be seen that when the number of columns is larger than 1300, the cost of Hankel matrix based \method{DeeP-LCC} and Page matrix based \method{DeeP-LCC} is stable around a constant. However, when the number of columns is less than 1300, compared to the Hankel matrix based \method{DeeP-LCC}, the Page matrix based ones show smaller cost fluctuation. This is due to the fact that the Page matrix can store more data than the Hankel matrix with the same number of columns, and hence contains more information about the system.
		
\section{Conclusion} \label{sec_conclusion}
This paper has presented a privacy-preserving \method{DeeP-LCC} for CAVs in a mixed-traffic environment. We have considered external eavesdropper and honest-but-curious central unit who intend to infer the CAVs' system states and inputs. A simple yet effective affine transformation mechanism has been designed to enable privacy preservation, and an extended form of the data-enabled predictive control has been derived to achieve safe and optimal control for CAVs. The proposed affine transformation mechanism can be seamlessly integrated into the data-enabled control without affecting the control performance. Theoretical and simulation results confirm that by using the proposed method, the leading cruise control \cite{Wang2021TITS} in mixed traffic can be addressed while avoiding disclosing private information to the attacker.

There are several interesting future directions: first, the current privacy-preserving \method{DeeP-LCC} follows a centralized control fashion that may not be scalable, and thus distributed versions are worth further investigation for large-scale mixed traffic systems; second, it is very interesting to study mixed traffic control when there exist communication delays and/or only a portion of HDV data is accessible.

%\balance
\appendices
\section{ Proof of Proposition \ref{Prop:Hankel_privacy}} \label{Appen:Prof_Hankel}
\begin{IEEEproof}
	If $ (\bar{u}_\mathrm{ini},\epsilon_\mathrm{ini},\bar{y}_\mathrm{ini},\bar{u},\epsilon,\bar{y})$ is a trajectory of \eqref{Eq:DT_TrafficModelAffine} with initial state being $x(t-T_{\mathrm{ini}})$, then the input-output response over $\left[ t-T_{\mathrm{ini}}, t+N-1 \right]$ can be expressed as
	\begin{equation} \label{Eq:InputOutput}
		%\bar{y}_{\left[ t-T_{\mathrm{ini}}, t+N-1 \right]} 
		\begin{bmatrix}
			\bar{y}_{\mathrm{ini}} \\ \bar{y}
		\end{bmatrix}
		= \begin{bmatrix}
			\mathcal{T}_{u} & \mathcal{T}_{\epsilon} & \mathcal{T}_{x} & \mathcal{T}_{l}
		\end{bmatrix}
		\begin{bmatrix}
			%\bar{u}_{\left[ t-T_{\mathrm{ini}}, t+N-1 \right] } \\
			\bar{u}_{\mathrm{ini}} \\ \bar{u} \\ \hline
			%\bar{\epsilon}_{\left[ t-T_{\mathrm{ini}}, t+N-1 \right] }
			\epsilon_{\mathrm{ini}} \\ \epsilon
			\\ \hline
			x(t-T_{\mathrm{ini}})
			\\ \hline
			1
		\end{bmatrix},
	\end{equation}
	%where $\bar{y}_{\left[ t-T_{\mathrm{ini}}, t+N-1 \right] } = \mathrm{col}(\bar{y}_{\mathrm{ini}}, \bar{y})$, $\bar{u}_{\left[ t-T_{\mathrm{ini}}, t+N-1 \right] } = \mathrm{col}(\bar{u}_{\mathrm{ini}}, \bar{u})$, and $\bar{\epsilon}_{\left[ t-T_{\mathrm{ini}}, t+N-1 \right] } = \mathrm{col}(\bar{\epsilon}_{\mathrm{ini}}, \bar{\epsilon})$. 
	where the matrices $\mathcal{T}_{u}$, $\mathcal{T}_{\epsilon}$, $\mathcal{T}_{x}$, $\mathcal{T}_{l}$ are given by
	\[
	\begin{aligned}
		\mathcal{T}_{u} & = \colvec{
			0 & 0 & 0 & \cdots & 0
			\\
			\bar{C}_{d}\bar{B}_{d} & 0 & 0 & \cdots & 0
			\\
			\bar{C}_{d}A_{d}\bar{B}_{d} & \bar{C}_{d}\bar{B}_{d} & 0 & \cdots & 0
			\\
			\vdots & \vdots & \vdots & \ddots & \vdots
			\\
			\bar{C}_{d}A_{d}^{T_{\mathrm{ini}}+N-2}\bar{B}_{d} & \bar{C}_{d}A_{d}^{T_{\mathrm{ini}}+N-3}\bar{B}_{d} &
			\bar{C}_{d}A_{d}^{T_{\mathrm{ini}}+N-4}\bar{B}_{d} & \cdots & 0
		},
		\\
		\mathcal{T}_{\epsilon} & = \colvec{
			0 & 0 & 0 & \cdots & 0
			\\
			\bar{C}_{d}H_{d} & 0 & 0 & \cdots & 0
			\\
			\bar{C}_{d}A_{d}\bar{B}_{d} & \bar{C}_{d}\bar{B}_{d} & 0 & \cdots & 0
			\\
			\vdots & \vdots & \vdots & \ddots & \vdots
			\\
			\bar{C}_{d}A_{d}^{T_{\mathrm{ini}}+N-2}H_{d} & \bar{C}_{d}A_{d}^{T_{\mathrm{ini}}+N-3}H_{d} & 
			\bar{C}_{d}A_{d}^{T_{\mathrm{ini}}+N-4}H_{d} &\cdots & 0
		},	\end{aligned}
	\]
	\[
	\begin{aligned}
		\mathcal{T}_{x} & = \begin{bmatrix}
			\bar{C}_{d} \\ \bar{C}_{d}A_{d} \\ \vdots \\ \bar{C}_{d}A_{d}^{T_{\mathrm{ini}}+N-1}
		\end{bmatrix}, \mathcal{T}_{l} = \begin{bmatrix}
			L_{y} \\ L_{y}+\bar{C}_{d}\bar{L}_{u} \\ \vdots \\ L_{y}+ \sum_{i=0}^{i=T_{\mathrm{ini}}+N-2}\!\bar{C}_{d}A_{d}^{i}\bar{L}_{u}
		\end{bmatrix}.
	\end{aligned}
	\]
	%Denote $x^{\mathrm{d}}_{\left[ 1, T-T_{\mathrm{ini}-N+1}\right]} = \mathrm{col}(x^{\mathrm{d}}(1),\ldots,x^{\mathrm{d}}(T-T_{\mathrm{ini}-N+1}))$ as the state sequence of length $T-T_{\mathrm{ini}-N+1}$ with respect to the pre-collected system trajectory $\left( \bar{u}^{\mathrm{d}}, \epsilon^{\mathrm{d}}, \bar{y}^{\mathrm{d}} \right)$.
	Considering that $\left( \bar{u}^{\mathrm{d}}, \epsilon^{\mathrm{d}}, \bar{y}^{\mathrm{d}} \right)$ is a pre-collected trajectory of \eqref{Eq:DT_TrafficModelAffine} and the definition of $\left( \bar{Y}_{p}^{\mathscr{H}}, \bar{Y}_{f}^{\mathscr{H}}, \bar{U}_{p}^{\mathscr{H}}, \bar{U}_{f}^{\mathscr{H}}, E_{p}^{\mathscr{H}}, E_{f}^{\mathscr{H}} \right)$, we derive  that
	\begin{equation} \label{Eq:InputOutputHankle}
		\begin{bmatrix}
			\bar{Y}_{p}^{\mathscr{H}} \\ \bar{Y}_{f}^{\mathscr{H}}
		\end{bmatrix} = \begin{bmatrix}
			\mathcal{T}_{u} & \mathcal{T}_{\epsilon} & \mathcal{T}_{x} & \mathcal{T}_{l}
		\end{bmatrix}
		\begin{bmatrix}
			\bar{U}_{p}^{\mathscr{H}} \\ \bar{U}_{f}^{\mathscr{H}} \\ \hline
			E_{p}^{\mathscr{H}} \\ E_{f}^{\mathscr{H}} \\ \hline
			X_{\mathrm{d}}^{\mathscr{H}}
			%\begin{bmatrix}
			%	x^{\mathrm{d}}(1), \ldots, x^{\mathrm{d}}(T-T_{\mathrm{ini}-N+1})
			%\end{bmatrix}
			%(x^{\mathrm{d}}_{\left[ 1, T-T_{\mathrm{ini}-N+1}\right]})^{\top}
			\\ \hline
			1_{T-T_{\mathrm{ini}}-N+1}^{\top}
		\end{bmatrix},
	\end{equation}
	where $X_{\mathrm{d}}^{\mathscr{H}} = \begin{bmatrix}
		x^{\mathrm{d}}(1), x^{\mathrm{d}}(2), \ldots, x^{\mathrm{d}}(T-T_{\mathrm{ini}}-N+1)
	\end{bmatrix}\in \mathbb{R}^{2n\times (T-T_{\mathrm{ini}}-N+1)}$.

We next show that the following matrix
\begin{equation} \label{Eq:Rank}
\begin{bmatrix}
	\bar{U}_{p}^{\mathscr{H}} \\ \bar{U}_{f}^{\mathscr{H}} \\ \hline
	E_{p}^{\mathscr{H}} \\ E_{f}^{\mathscr{H}} \\ \hline
	X_{\mathrm{d}}^{\mathscr{H}}
	\\ \hline
	1_{T-T_{\mathrm{ini}}-N+1}^{\top}
\end{bmatrix}
\end{equation}
has full row rank. It is clear that~\eqref{Eq:DT_TrafficModelAffine} can be rewritten as
\begin{equation} \label{Eq:DT_TrafficModelAffine2}
	\begin{bmatrix}
		x(k+1) \\ 1
	\end{bmatrix} = \check{A}_{\mathrm{d}} \begin{bmatrix}
		x(k) \\ 1
	\end{bmatrix} + \check{B}_{\mathrm{d}} \begin{bmatrix}
		\bar{u}(k) \\ \epsilon(k)
	\end{bmatrix},
\end{equation}
where
\begin{equation} \label{Eq:check_A}
	\check{A}_{\mathrm{d}} = \begin{bmatrix}
		A_{\mathrm{d}} & \bar{L}_{u} \\
		0 & 1
	\end{bmatrix}, \check{B}_{\mathrm{d}} = \begin{bmatrix}
		\bar{B}_{\mathrm{d}} & H_{\mathrm{d}} \\
		0 & 0
	\end{bmatrix}.
\end{equation} 
Let $\begin{bmatrix}
	\eta_{u}, \eta_{\epsilon}, \xi_{x}, \xi_{l}
\end{bmatrix}$ be a vector in the left kernel of \eqref{Eq:Rank}, where $\eta_{u}^{\top}\in \mathbb{R}^{m(T_{\mathrm{ini}}+N)}$, $\eta_{\epsilon}^{\top}\in \mathbb{R}^{T_{\mathrm{ini}}+N}$, $\xi_{x}^{\top} \in \mathbb{R}^{2n}$, and $\xi_{l}\in \mathbb{R}$. Based on~\eqref{Eq:Rank}, \eqref{Eq:DT_TrafficModelAffine2}, and the condition that $\hat{\bar{u}}^\mathrm{d}_{\left[1, T\right]}$, i.e., $\left( \bar{u}^{\mathrm{d}}_{\left[1, T\right]}, \epsilon^{\mathrm{d}}_{\left[1, T\right]} \right)$, is Hankel exciting of order $T_{\mathrm{ini}}+N+2n+1$, we can follow the same arguments from the proof in~\cite[Theorem 1]{Van2020CSL} to obtain that $\eta_{u} = 0$, $\eta_{\epsilon}=0$, and
\begin{equation} \label{Eq:xi}
\begin{bmatrix}
	\xi_{x}, \xi_{l}
\end{bmatrix}\check{B}_{\mathrm{d}}=\begin{bmatrix}
\xi_{x}, \xi_{l}
\end{bmatrix}\check{A}_{\mathrm{d}}\check{B}_{\mathrm{d}} = \cdots = \begin{bmatrix}
\xi_{x}, \xi_{l}
\end{bmatrix}\check{A}_{\mathrm{d}}^{2n}\check{B}_{\mathrm{d}} = 0.
\end{equation}
According to~\eqref{Eq:check_A} and \eqref{Eq:xi}, it can be derived that
\begin{equation} \label{Eq:xi_x}
	\xi_{x} \begin{bmatrix}
		\bar{B}_{\mathrm{d}} \!\!&\!\! H_{\mathrm{d}}
	\end{bmatrix} = \xi_{x} A_{\mathrm{d}} \begin{bmatrix}
	    \bar{B}_{\mathrm{d}} \!\!&\!\! H_{\mathrm{d}}
    \end{bmatrix} = \!\cdots\! = \xi_{x} A_{\mathrm{d}}^{2n} \begin{bmatrix}
    \bar{B}_{\mathrm{d}} \!\!&\!\! H_{\mathrm{d}}
\end{bmatrix} = 0.
\end{equation}
Since $\left(A_{\mathrm{d}}, \begin{bmatrix}
	B_{\mathrm{d}} \!\!&\!\! H_{\mathrm{d}}
\end{bmatrix} \right)$ is controllable and $\bar{B}_{\mathrm{d}}=B_{\mathrm{d}}P_{u}^{-1}$, we can conclude that $\left(A_{\mathrm{d}}, \begin{bmatrix}
\bar{B}_{\mathrm{d}} \!\!&\!\! H_{\mathrm{d}}
\end{bmatrix} \right)$ is controllable. \eqref{Eq:xi_x} and the controllability of $\left(A_{\mathrm{d}}, \begin{bmatrix}
\bar{B}_{\mathrm{d}} \!\!&\!\! H_{\mathrm{d}}
\end{bmatrix} \right)$ reveal that $\xi_{x}=0$. Furthermore, as $\begin{bmatrix}
\eta_{u}, \eta_{\epsilon}, \xi_{x}, \xi_{l}
\end{bmatrix}$ is a vector in the left kernel of~\eqref{Eq:Rank}, $\eta_{u} = 0$, $\eta_{\epsilon}=0$, and $\xi_{x}=0$, we have $\xi_{l} 1_{T-T_{\mathrm{ini}}-N+1}^{\top} = 0$, indicating that $\xi_{l}=0$. Based on $\begin{bmatrix}
\eta_{u}, \eta_{\epsilon}, \xi_{x}, \xi_{l}
\end{bmatrix}=0$, we can conclude that \eqref{Eq:Rank} has full row rank.

    \iffalse
	By Theorem 1 in \cite{Van2020CSL}, the condition that $\hat{\bar{u}}^\mathrm{d}_{\left[1, T\right]}$, i.e., $\left( \bar{u}^{\mathrm{d}}_{\left[1, T\right]}, \epsilon^{\mathrm{d}}_{\left[1, T\right]} \right)$, is Hankel exciting of order $T_{\mathrm{ini}}+N+2n+1$ implies the full row rank property
	\begin{equation} \label{Eq:Rank}
		\mathrm{rank} \begin{bmatrix}
			\bar{U}_{p}^{\mathscr{H}} \\ \bar{U}_{f}^{\mathscr{H}} \\ \hline
			E_{p}^{\mathscr{H}} \\ E_{f}^{\mathscr{H}} \\ \hline
			X_{\mathrm{d}}^{\mathscr{H}}
			%\begin{bmatrix}
			%	x^{\mathrm{d}}(1), \ldots, x^{\mathrm{d}}(T-T_{\mathrm{ini}-N+1})
			%\end{bmatrix}
			%(x^{\mathrm{d}}_{\left[ 1, T-T_{\mathrm{ini}-N+1}\right]})^{\top}
			\\ \hline
			1_{T-T_{\mathrm{ini}}-N+1}^{\top}
		\end{bmatrix} = (m+1)(T_{\mathrm{ini}}+N)+2n+1.
	\end{equation} 
    \fi
	Based on the Rouch\'{e}–Capelli theorem, the full row rank of \eqref{Eq:Rank} implies the existence of $\bar{g}\in \mathbb{R}^{T-T_\mathrm{ini}-N+1}$ such that
	\begin{equation} \label{Eq:Umap}
		\begin{bmatrix}
			%\bar{u}_{\left[ t-T_{\mathrm{ini}}, t+N-1 \right] } \\
			\bar{u}_{\mathrm{ini}} \\ \bar{u} \\ \hline
			%\bar{\epsilon}_{\left[ t-T_{\mathrm{ini}}, t+N-1 \right] }
			\epsilon_{\mathrm{ini}} \\ \epsilon
			\\ \hline
			x(t-T_{\mathrm{ini}})
			\\ \hline
			1
		\end{bmatrix} = 
		\begin{bmatrix}
			\bar{U}_{p}^{\mathscr{H}} \\ \bar{U}_{f}^{\mathscr{H}} \\ \hline
			E_{p}^{\mathscr{H}} \\ E_{f}^{\mathscr{H}} \\ \hline
			X_{\mathrm{d}}^{\mathscr{H}}
			%\begin{bmatrix}
			%	x^{\mathrm{d}}(1), \ldots, x^{\mathrm{d}}(T-T_{\mathrm{ini}-N+1})
			%\end{bmatrix}
			%(x^{\mathrm{d}}_{\left[ 1, T-T_{\mathrm{ini}-N+1}\right]})^{\top}
			\\ \hline
			1_{T-T_{\mathrm{ini}}-N+1}^{\top}
		\end{bmatrix}\bar{g}.
	\end{equation}
	Substituting \eqref{Eq:InputOutputHankle} and \eqref{Eq:Umap} into \eqref{Eq:InputOutput}, it can be obtained that
	\begin{equation} \label{Eq:Ymap}
		\begin{bmatrix}
			\bar{y}_{\mathrm{ini}} \\ \bar{y}
		\end{bmatrix} = \begin{bmatrix}
			\bar{Y}_{p}^{\mathscr{H}} \\ \bar{Y}_{f}^{\mathscr{H}}
		\end{bmatrix}\bar{g}.
	\end{equation}
	From \eqref{Eq:Umap} and \eqref{Eq:Ymap}, it can be concluded that there exists $\bar{g}$ such that \eqref{Eq:AdaptedDeePCAchievabilityAffine_privacy} is satisfied, which completes the proof.
\end{IEEEproof}
		
\section{Proof of Proposition \ref{Prop:Page_privacy}} \label{Appen:Prof_Page}
\begin{IEEEproof}
	According to the system dynamics and the definition of $\left( \bar{Y}_{p}^{\mathscr{P}}, \bar{Y}_{f}^{\mathscr{P}}, \bar{U}_{p}^{\mathscr{P}}, \bar{U}_{f}^{\mathscr{P}}, E_{p}^{\mathscr{P}}, E_{f}^{\mathscr{P}} \right)$, we have
	\begin{equation} \label{Eq:InputOutputPage}
		\begin{bmatrix}
			\bar{Y}_{p}^{\mathscr{P}} \\ \bar{Y}_{f}^{\mathscr{P}}
		\end{bmatrix} = \begin{bmatrix}
			\mathcal{T}_{u} & \mathcal{T}_{\epsilon} & \mathcal{T}_{x} & \mathcal{T}_{l}
		\end{bmatrix}
		\begin{bmatrix}
			\bar{U}_{p}^{\mathscr{P}} \\ \bar{U}_{f}^{\mathscr{P}} \\ \hline
			E_{p}^{\mathscr{P}} \\ E_{f}^{\mathscr{P}} \\ \hline
			X_{\mathrm{d}}^{\mathscr{P}}
			%\begin{bmatrix}
			%	x^{\mathrm{d}}(1), \ldots, x^{\mathrm{d}}(T-T_{\mathrm{ini}-N+1})
			%\end{bmatrix}
			%(x^{\mathrm{d}}_{\left[ 1, T-T_{\mathrm{ini}-N+1}\right]})^{\top}
			\\ \hline
			1_{\lfloor \frac{T}{T_{\mathrm{ini}+N}} \rfloor}^{\top}
		\end{bmatrix},
	\end{equation}
	where 
	\[
	\begin{aligned}
		&X_{\mathrm{d}}^{\mathscr{P}} = \left[
		x^{\mathrm{d}}(1), x^{\mathrm{d}}(T_{\mathrm{ini}}+N+1), \ldots, \right.
		\\
		&\qquad \left. x^{\mathrm{d}}( (\lfloor \tfrac{T}{T_{\mathrm{ini}}+N} \rfloor -1)(T_{\mathrm{ini}}+N)+1)
		\right]\in \mathbb{R}^{2n\times ( \lfloor \tfrac{T}{T_{\mathrm{ini}}+N} \rfloor )}.
	\end{aligned}
	\]
	By following the similar arguments from Theorem 2.1 in \cite{Coulson2021TAC} and from Appendix A, it can be proved that if $\hat{\bar{u}}^\mathrm{d}_{\left[1, T\right]}$, i.e., $\left( \bar{u}^{\mathrm{d}}_{\left[1, T\right]}, \epsilon^{\mathrm{d}}_{\left[1, T\right]} \right)$, is $(T_{\mathrm{ini}}+N)$-Page exciting of order $2n+2$, then the following full row rank property is satisfied:
	\begin{equation} \label{Eq:Rank_Page}
		\mathrm{rank} \begin{bmatrix}
			\bar{U}_{p}^{\mathscr{P}} \\ \bar{U}_{f}^{\mathscr{P}} \\ \hline
			E_{p}^{\mathscr{P}} \\ E_{f}^{\mathscr{P}} \\ \hline
			X_{\mathrm{d}}^{\mathscr{P}}
			%\begin{bmatrix}
			%	x^{\mathrm{d}}(1), \ldots, x^{\mathrm{d}}(T-T_{\mathrm{ini}-N+1})
			%\end{bmatrix}
			%(x^{\mathrm{d}}_{\left[ 1, T-T_{\mathrm{ini}-N+1}\right]})^{\top}
			\\ \hline
			1_{\lfloor \frac{T}{T_{\mathrm{ini}+N}} \rfloor}^{\top}
		\end{bmatrix} = (m+1)(T_{\mathrm{ini}}+N)+2n+1.
	\end{equation}   
	
 Based on \eqref{Eq:InputOutput}, \eqref{Eq:InputOutputPage} and the rank condition \eqref{Eq:Rank_Page}, it can be concluded that there exists a vector $\bar{g}\in \mathbb{R}^{\lfloor \frac{T}{T_{\mathrm{ini}}+N} \rfloor}$ such that
	\begin{equation} \label{Eq:Umap_Page}
		\begin{bmatrix}
			%\bar{u}_{\left[ t-T_{\mathrm{ini}}, t+N-1 \right] } \\
			\bar{u}_{\mathrm{ini}} \\ \bar{u} \\ \hline
			%\bar{\epsilon}_{\left[ t-T_{\mathrm{ini}}, t+N-1 \right] }
			\epsilon_{\mathrm{ini}} \\ \epsilon
			\\ \hline
			x(t-T_{\mathrm{ini}})
			\\ \hline
			1
		\end{bmatrix} = 
		\begin{bmatrix}
			\bar{U}_{p}^{\mathscr{P}} \\ \bar{U}_{f}^{\mathscr{P}} \\ \hline
			E_{p}^{\mathscr{P}} \\ E_{f}^{\mathscr{P}} \\ \hline
			X_{\mathrm{d}}^{\mathscr{P}}
			%\begin{bmatrix}
			%	x^{\mathrm{d}}(1), \ldots, x^{\mathrm{d}}(T-T_{\mathrm{ini}-N+1})
			%\end{bmatrix}
			%(x^{\mathrm{d}}_{\left[ 1, T-T_{\mathrm{ini}-N+1}\right]})^{\top}
			\\ \hline
			1_{\lfloor \frac{T}{T_{\mathrm{ini}+N}} \rfloor}^{\top}
		\end{bmatrix}\bar{g},
	\end{equation}
	and 
	\begin{equation} \label{Eq:Ymap_Page}
		\begin{bmatrix}
			\bar{y}_{\mathrm{ini}} \\ \bar{y}
		\end{bmatrix} = \begin{bmatrix}
			\bar{Y}_{p}^{\mathscr{P}} \\ \bar{Y}_{f}^{\mathscr{P}}
		\end{bmatrix}\bar{g}.
	\end{equation}
	Finally, we can conclude from \eqref{Eq:Umap_Page} and \eqref{Eq:Ymap_Page} that there exists $\bar{g}$ such that \eqref{Eq:PageDeePCAchievabilityAffine_privacy} is satisfied.
\end{IEEEproof}

\balance
\bibliographystyle{IEEEtran}
\bibliography{IEEEabrv,reference}
\end{document}